%% file: document.tex
\documentclass[11pt]{article}
\usepackage[pdfborder={0 0 0 0}]{hyperref}
\include{commands}
\newcommand{\as}[2]{_{\{ #1 \} #2 }}

\begin{document}

\title{Correlation functions for Schr\"odinger backgrounds}
\author{Balt C. van Rees\thanks{vanrees@insti.physics.sunysb.edu}}
\date{}
\maketitle

\begin{center}
\it C. N. Yang Institute for Theoretical Physics\\
State University of New York, Stony Brook, NY 11794-3840
\end{center}

\vskip 1cm 
\begin{abstract}
We work out the holographic dictionary for the three-dimensional Schr\"odinger spacetimes. The first step in our analysis involves the correct identification of the dual sources from the radial expansion of the bulk fields, which turns out to be surprisingly subtle. We discuss in detail the holographic renormalization procedure at the linearized level and holographically compute the two-point functions of the energy-momentum tensor and an irrelevant vector operator. We discuss the appearance of multi-trace counterterms, parametrize the scheme dependence in our results, identify the non-relativistic Ward identities and compare them with expectations in the literature. Our results lead to valuable general insights regarding holography for spacetimes that are not of an asymptotically AdS form.
\end{abstract}

\newpage
\tableofcontents
\newpage

\section{Introduction}
In this paper we perform a detailed holographic analysis for the three-dimensional version of the `Schr\"odinger' spacetimes of \cite{Son:2008ye,Balasubramanian:2008dm} with dynamical exponent $z=2$, using three-dimensional Einstein gravity coupled to a free massive vector field $A_\mu$ as the effective theory in the bulk. The metric and vector field of the background spacetime are given by:
\be
\label{eq:schrbg}
ds^2 = dr^2 - b^2 e^{4r} du^2 + 2 e^{2r} du dv \qquad \qquad A_\mu dx^\mu = b e^{2r} du
\ee  
with $b$ a real parameter. For nonzero $b$ this spacetime cannot be conformally compactified in the usual sense \cite{FeffermanGraham,Graham:1999jg} and therefore it is not an Asymptotically locally AdS spacetime. Correspondingly, the dual theory is not expected to be conformal in the UV. Instead, significant evidence has been obtained \cite{Son:2008ye,us,Costa:2010cn,Kraus:2011pf} to support the fact that the dual field theory is given by a \emph{finite irrelevant} deformation of the two-dimensional CFT which is dual to the AdS$_3$ solution given by \eqref{eq:schrbg} with $b = 0$. Finite irrelevant deformations of field theories generally lead to non-renormalizability and infinitely many counterterms and therefore result in a certain amount of non-locality and also a significant scheme dependence in physical observables. This non-locality and scheme dependence will be found in the holographic analysis as well. Our analysis will ultimately lead to explicit expressions for (the scheme-independent part of) the two-point functions of the energy-momentum tensor $T_{ij}$ and of the vector operator $ \op_i$ dual to $A_\mu$. We present these two-point functions in section \ref{subsec:twoptfiniteb} below. The results of this paper also remove any conceptual difficulties in extending the dictionary to higher-point correlation functions, although technically the computations may become very involved.

Our motivations for performing this analysis are the following. First, the fact that Schr\"odinger spacetime is `close' to AdS, in the sense that we can smoothly dial $b$ back to zero and obtain precisely an empty AdS spacetime, offers us a unique opportunity to study in a controlled way the field theory dual to a spacetime that is not of an Asymptotically locally AdS (or AlAdS) form. The holographic analysis of such non-AlAdS spacetimes is in many ways still in its infancy and the results obtained in this paper should be very useful for further investigations in this direction. In fact, the gradual loss of locality resulting from an irrelevant deformation is very similar what happens if one gradually lowers a field theory cutoff to finite values (\ie if one works in inverse powers of $\Lambda$) and our general results should therefore also be of relevance for a possible implementation of the Wilsonian renormalization group flow in the bulk theory \cite{Heemskerk:2010hk,Faulkner:2010jy,Brattan:2011my}.

Another motivation for our work are several discussions in the literature surrounding in particular the Schr\"odinger spacetimes with $z = 2$ \cite{Compere:2009qm,Anninos:2010pm,Guica:2011ia,ElShowk:2011cm,Song:2011sr,Kraus:2011pf,us} or the symmetry structure of the dual field theory \cite{Hofman:2011zj}. We can add the following results to this discussion. In agreement with general expectations, we will find below that the non-relativistic deformation preserves a single (`left-moving') Virasoro algebra as well as translation invariance in the other (`right-moving') direction, at least at the level of the two-point functions. The left-moving Virasoro algebra includes the left-moving (`chiral') dilatation current which confirms the results of \cite{Kraus:2011pf,us} regarding the exact marginality of the deformation from a non-relativistic viewpoint. On the other hand, we do not find any enhancements to an infinite-dimensional symmetry algebra in the right-moving sector. This contradicts the expectations of \cite{Anninos:2010pm} and, in a sense, also of \cite{Hofman:2011zj}. More details will be presented below.

Let us briefly discuss some existing results regarding holography for non-AlAdS spacetimes. Many of these results are valid for spacetimes that are conformally related to AdS spacetimes, see for example \cite{Wiseman:2008qa,Kanitscheider:2008kd,Kanitscheider:2009as,Gouteraux:2011qh,Papadimitriou:2011qb} for geometries dual to non-conformal branes, or \cite{Aharony:2005zr,Borodatchenkova:2008fw} for results involving the cascading theories. See also \cite{Papadimitriou:2010as} for a more abstract approach. Furthermore, several aspects of holography for Schr\"odinger spacetimes have already been discussed in \cite{us} where in particular the `irrelevant deformation' viewpoint of the spacetime was already worked out in considerable detail. Some very interesting results have also been obtained \cite{Ross:2009ar,Ross:2011gu,Baggio:2011cp,Mann:2011hg,Griffin:2011xs,Baggio:2011ha} for the `Lifshitz' solutions of \cite{Kachru:2008yh,Taylor:2008tg}. These spacetimes are again not AlAdS and their holographic analysis is complicated by very similar issues as those described here. So far, the general approach for the Lifshitz spacetimes has been to impose certain boundary conditions on the bulk fields and renormalize the on-shell action for solutions satisfying those boundary conditions. (Related approaches for Schr\"odinger spacetimes can be found in \cite{Hartong:2010ec,Compere:2009qm}.) In contrast, our approach is to renormalize the on-shell action for \emph{any} solution satisfying the equations of motion. This allows us to compute finite correlation functions of \emph{all} the boundary operators rather than just a subset of them, but it necessarily requires a perturbative approach since in the presence of an irrelevant deformation the asymptotics become increasingly involved at every order in perturbation theory. We however regard this as an unavoidable consequence of the irrelevant deformation. It would be very interesting to see if our methods can be extended to the Lifshitz spacetimes as well.

The Schr\"odinger solutions with five bulk dimensions were shown to be solutions of type IIB supergravity in \cite{Adams:2008wt,Herzog:2008wg,Maldacena:2008wh} with a general five-dimensional Sasaki-Einstein manifold as the compact part of the geometry. Furthermore, in \cite{Maldacena:2008wh} it was shown that one may consistently truncate the five-dimensional field content to only the five-dimensional metric, the massive vector field sourcing the deformation and three additional scalars. The field theory dual to these spacetimes was argued to be a so-called null dipole theory, described earlier in \cite{Bergman:2001rw}. The three-dimensional Schr\"odinger solutions of the form considered in this paper were also shown to be solutions of type IIB supergravity in \cite{Kraus:2011pf,Bobev:2011qx}, where supersymmetric solutions were found as well. The general structure of these solutions is very similar to the five-dimensional case and one may reasonably expect that the dual field theory is again a certain (two-dimensional) null dipole theory. We are however not aware of a consistent truncation to three bulk dimensions involving only the fields considered in this paper (\ie three-dimensional Einstein gravity coupled to a free massive vector field). Although this implies that our results cannot be directly interpreted as a valid string theory computation (yet), we do not expect the correlation functions obtained from for example a `null dipole deformed D1/D5 CFT' to be very different from those presented below. It would of course be very interesting to investigate this further.

\subsection{Approach}
The main obstacle in the holographic analysis of the background \eqref{eq:schrbg} is the unfamiliar asymptotic behavior of the fields near the boundary $r \to \infty$ for nonzero $b$. As we will see in section \ref{sec:asymptsol}, this unfamiliar behavior is exacerbated when we consider linearized fluctuations on top of the background \eqref{eq:schrbg} and leads to radial expansions of the bulk fields which are much more complicated than in the AdS case. On the other hand, all our solutions turn out to be analytic in $b$, so when we dial $b$ back to zero we can smoothly return to the linear fluctuations around the AdS background. This simple observation is the key to understanding holography for Schr\"odinger spacetimes. Namely, when we dial $b$ back up from zero, the perturbative expansion in $b$ is nothing more than an expansion in an irrelevant deformation parameter in the dual field theory. The holographic analysis of perturbatively small irrelevant deformations was worked out in detail in \cite{balt,baltlogs} and we can use the results of those papers to find the dual correlation functions up to any finite order in $b$. In this paper we perform a detailed holographic analysis up to order $b^4$ and use these results to conjecture the result of the asymptotic analysis to all orders in $b$. Our conjecture passes several consistency tests and could only be incorrect if there would be a significant structural change appearing at order $b^5$, a scenario we consider highly unlikely. Combining our knowledge of the linearized bulk solution to all orders in $b$ with our conjecture for the all-order results of the asymptotic analysis then allows us to extract the two-point functions of the dual field theory for finite $b$.

\subsection{Outline}
In the next section we set up the bulk problem by introducing the bulk Lagrangian, computing its equations of motion and discussing the linearized analysis. We also explain in more detail how to treat the problem perturbatively in $b$. We then holographically compute the relevant two-point functions for $b=0$, so for empty AdS$_3$, in section \ref{sec:linb0}. This gives us the basic correlators that will receive an infinite series of $b$-dependent corrections once we switch on $b$. In section \ref{sec:asymptsol} we present the solution to the linearized equations of motion for finite values of $b$. We explain how a careful $b \to 0$ limit leads to a proper identification of the sources in the radial expansion of the fields. The analysis in this section, in particular subsection \ref{subsec:interpret}, contains the most important conceptual results of this paper. We then continue with the holographic analysis: first to order $b^1$ in section \ref{sec:linb1} and then to order $b^4$ in section \ref{sec:linb4}. In fact, in section \ref{sec:linb4} we also present our conjecture for the all-order result and compute the associated two-point functions. In the appendix to this paper we briefly discuss the irrelevant deformation from a field theory perspective.  

\section{Setup}
\label{sec:setup}
We consider a three-dimensional Einstein-Proca bulk theory with the following action:
\be
S = \frac{1}{16\pi G_N} \int d^3 x \sqrt{-G} (R - 2 \Lambda - \qt F_{\m \n} F^{\m \n} - \hf m^2 A^\m A_\m) + \frac{1}{8 \pi G_N}  \int d^2 x \sqrt{-\g} K
\ee
and we set $\Lambda = -1$ henceforth. We can also set $16 \pi G_N = 1$ and we need $m^2 = 4$ in order for the Schr\"odinger spacetime with $z=2$ to be a solution. The equations of motion take the form:
\be
\label{eq:fulleom}
\begin{split}
&R_{\m \n} - \hf R G_{\m \n} - G_{\m \n} = \hf F_{\m \r} F\du{\n}{\r}+ 2 A_\m A_\n - G_{\m \n} (\frac{1}{8} F_{\r \s}F^{\r \s} +  A_\r A^\r )\\
&\cdel^\n F_{\n \m} - 4 A_\m = 0
\end{split}
\ee
We will always be working in a radial-axial gauge where:
\be
G_{\m \n} dx^\m dx^\n = dr^2 + \g_{ij} dx^i dx^j\,.
\ee
In this coordinate system the extrinsic curvature of a slice of constant $r$ is
\be
K_{ij} = \hf  \dot \g_{ij}\,,
\ee
where a dot denotes a radial derivative.

One solution to the equations of motion is empty AdS$_3$ with a metric and vanishing vector field:
\be
\label{eq:ads3bg}
\g_{ij}dx^i dx^j = 2 e^{2r} du dv \qquad \qquad A_\m = 0\,.
\ee
Another solution to the equations of motion is the Schr\"odinger spacetime:
\be
\g_{ij}dx^i dx^j = - b^2 e^{4r} du^2 + 2 e^{2r} du dv \qquad \qquad A_\m dx^\m = b e^{2r} du\,.
\ee
Notice that we work in conventions where $\eta_{ij} dx^i dx^j = 2 du dv$ so $\eta_{uv} = 1$ and lightcone indices can be raised and lowered without introducing factors of $2$. 

When we attempt to compute the on-shell action for the above solutions we find divergences as we integrate all the way to the conformal boundary at $r = \infty$. We cancel these in the usual fashion by adding a counterterm action. In our case we add counterterms of the form:
\be
\label{eq:Sctpostulated}
S_{\text{ct}} = \int d^2 x \sqrt{- \g} (- 2 + A_i A^i)\,,
\ee
which ensures that the on-shell action remains finite for both of the above solutions. Furthermore, the first variation of the renormalized on-shell action
\be
\label{eq:deltasonshellbg}
\delta (S + S_{\text{ct}}) = \int d^2 x \sqrt{-\g} (- K^{ij} + \g^{ij}(- 1 + K + \hf A_k A^k) - A^i A^j) \delta \g_{ij} + \int d^2 x \sqrt{-\g} (F^{i r} + 2 A^i) \delta A_i
\ee
also vanishes on the above backgrounds with this choice of counterterms. We note that this `ad hoc' definition of a counterterm action is somewhat illegal, since in principle one should find the most general asymptotic solution to the equations of motion (to the desired order in the fluctuations) and then one \emph{computes} rather than postulates the counterterm action by following the usual holographic renormalization methods. However this is precisely what we will do below when we compute the counterterm action for perturbations on top of these backgrounds. Therefore, if our initial guess \eqref{eq:Sctpostulated} for the counterterm action happened to be wrong then we will automatically compute, order by order in the fluctuations, the corrections that bring us to the right counterterm action. Of course, our reason for adding \eqref{eq:Sctpostulated} is that it provides us with a convenient starting point.

To find the two-point functions in the dual field theory we will below perform a linearized analysis of small fluctuations around these backgrounds. For the metric variations we stay in the radial-axial gauge so $\delta G_{rr} = 0$, $\delta G_{r i} = 0$ and we define $\delta \g_{ij} = h_{ij}$. For the vector perturbations we define $\delta A_\m = a_\m$ and $\delta F_{\m \n} = f_{\m \n}$. Henceforth $\g_{ij}$ and $A_i$ will denote the fixed background metric and gauge field, respectively, and indices are raised with $\g^{ij}$. In summary, then, we have:
\be
\g_{ij} \to \g_{ij} + h_{ij} \qquad \qquad A_\mu \to A_\mu + a_\mu\,.
\ee
We will need to compute the on-shell action to second order in the fluctuations. Expanding \eqref{eq:deltasonshellbg} we find that this on-shell action takes the form:
\be
\label{eq:sonshell2ndorder}
S = \int d^2 x \sqrt{-\g} (h^{ij} - \hf \dot h^{ij} + \g^{ij}(\hf \dot h^k_k + A^k a_k + \hf A_k h^{kl} A_l) - 2 A^i a^j) h_{ij} + \int d^2 x \sqrt{-\g} (f^{i r} + 2 a^i) a_i
\ee
where $\dot h^{ij}$ is shorthand for $\g^{ik} \dot h_{kl} \g^{lj}$ and similarly $\dot h^k_k = \g^{kl} \dot h_{kl}$. Notice that the bulk term in the on-shell action for the fluctuations vanishes because both the background and the fluctuations satisfy the equations of motion.

As we announced in the introduction to this paper, the main idea of the subsequent computations will be to work perturbatively in the deformation parameter $b$. In this sense $b$ is exactly like the coupling constant $\lambda$ of the $\Phi^3$ model considered in \cite{balt}. In particular, the solution to the linearized equations of motion will have an expansion in $b$,
\be
\begin{split}
a_i &= a_{\{0\}i} + a_{\{1\}i} + a_{\{2\}i} + \ldots\\
h_{ij} &= h_{\{0\}ij} + h_{\{1\}ij} + h_{\{2\}ij} + \ldots 
\end{split} 
\ee
where the curly brackets indicate the corresponding order in $b$. Furthermore, the background fields $\g_{ij}$ and $A_i$, which appear in the linearization of the equations of motion \eqref{eq:fulleom} and in the on-shell action \eqref{eq:sonshell2ndorder}, also have an expansion in $b$. These expansions however terminate at order $b^2$. Again this is just as in the $\lambda \Phi^3$ model of \cite{balt} where the on-shell action also contains an explicit $\lambda \Phi^3$ term.
 
\section{Analysis for $b=0$}
\label{sec:linb0}
Our first task is to perform the holographic analysis of linearized fluctuations around empty AdS$_3$. We henceforth work in lightcone momentum space with lightcone momenta $q_u$ and $q_v$. The analysis in this section closely follows standard results in the holographic renormalization literature, see in particular \cite{deHaro:2000xn,Skenderis:1999nb,Bianchi:2001kw}. We refer to \cite{Skenderis:2002wp} for an introduction to holographic renormalization.

\subsection{Linearized asymptotic solution}
The linearization of the equations of motion \eqref{eq:fulleom} around the AdS$_3$ background \eqref{eq:ads3bg} is straightforward. One finds that the metric and vector field fluctuations decouple and the equations are easily solved. For the metric we find the following solution:
\be
h_{ij} = e^{2r} h_{(0)ij} + h_{(2)ij} 
\ee
where the integration constants $h_{(0)ij}$ are the sources and
\be
\begin{split}
&h_{(2)i}^i = - \frac{1}{2} (\del^i \del^j  h_{(0)ij} - \del^i \del_j h_{(0)i}^i)\\
&\del_j h_{(2)i}^i - \del^i h_{(2)ij} = 0\,.
\end{split}
\ee
Below, we will be working exclusively in momentum space and in components. In that language the above equations become:
\be
\label{eq:h2constraints}
\begin{split}
&h_{(2)uv} = \qt(q_v^2 h_{(0)uu} - 2 q_u q_v h_{(0)uv} + q_u^2 h_{(0)vv})\\
&q_v h_{(2)uv} - q_u h_{(2)vv} = 0\\
&q_u h_{(2)uv} - q_v h_{(2)uu} = 0\,.
\end{split}
\ee
As expected for Einstein gravity in AdS$_3$, this solution for $h_{ij}$ is just a linearized diffeomorphism.

For the massive vector field we obtain the following asymptotic solution to the linearized equation of motion: 
\be
\label{eq:aexpansion}
a_i = a_{(0)i} e^{2r} + a_{(2)i} + (r \tilde a_{(4)i} + a_{(4)i}) e^{-2r} + \ldots 
\ee
where $a_{(0)i}$ are the source terms and the first two subleading terms are given by:
\be
\begin{split}
a_{(2)u} &= \qt q_u (q_u a_{(0)v} - q_v a_{(0)u})\\
\tilde a_{(4)u} &= - \qt q_u^3 q_v a_{(0)v} 
\end{split}
\ee
and similar equations hold with $u$ and $v$ interchanged. As usual, the $a_{(4)i}$ are not determined by the asymptotic analysis of the equations of motion alone but will be fixed by demanding regularity in the interior of the spacetime.

\subsection{Holographic renormalization}
Upon substitution of the asymptotic solution into the on-shell action, we find divergences as we send the cutoff $r \to \infty$. These are cancelled by the counterterm action:
\be
S_{\text{ct}} = - \hf \int d^2 x \sqrt{\g} \Big( \hf a^k \square_\g a_k + (\del_k a^k)^2 + r (\qt a^k \square_\g^2 a_k + \hf \del^k a_k \square_\g \del^m  a_m)\Big)
\ee
with $\square_\gamma \equiv e^{-2r} \square \equiv e^{-2r} \delta^{ij} \del_i \del_j$ the Laplacian associated to the induced metric $\g_{ij}$. This counterterm action can be found with the usual means, in particular we obtained it using the Hamilton-Jacobi method of \cite{Papadimitriou:2004ap}. We will not repeat its derivation here. Notice that the counterterm action is covariant up to the explicit appearance of the coordinate $r$, which signals the presence of a conformal anomaly.   

\subsection{One-point functions}
After adding the counterterms, the first variation of the renormalized on-shell action becomes finite and takes the form:
\be
\delta (S + S_{\text{ct},0}) = \int d^2 x \Big[ (4 a_{(4)}^i - \frac{3}{16} \square^2 a_{(0)}^i + \frac{1}{8} \square \del^i \del^k a_{(0)k}) \delta a_{(0)i} + ( h_{(2)}^{ij} + \hf \tilde R[h_{(0)}] \eta^{ij}) \delta h_{(0)ij}  \Big]
\ee 
with $\tilde R[h_{(0)}] = \del^i \del^j h_{(0)ij} - \del^i \del_i h_{(0)j}^j$ the linearized curvature associated to $h_{(0)ij}$. The one-point functions are then given by:
\be
\label{eq:oneptfns}
\begin{split}
\vev{\op^i} &= - \frac{1}{\sqrt{- g_{(0)}}} \fdel{a_{(0)i}} (S + S_{\text{ct},0})  = - 4 a_{(4)}^i + \frac{3}{16} \square^2 a_{(0)}^i - \frac{1}{8} \square \del^i \del^k a_{(0)k} \\
\vev{T^{ij}} &= \frac{4\pi}{\sqrt{- g_{(0)}}} \fdel{h_{(0)ij}} (S + S_{\text{ct},0}) = 4 \pi (h_{(2)}^{ij} + \hf \tilde R[h_{(0)}] \eta^{ij} )\,.
\end{split}
\ee
These are one-point functions in the presence of sources, so differentiating them once more with respect to the sources gives the two-point functions, see below. The contact terms in $\vev{\op^i}$ are scheme-dependent, although conformal invariance constrains them to a linear combination of $\del^i \square \del^j a_{(0)j}$ and $\square^2 a_{(0)}^i$. We will elaborate on the scheme dependence of our results in the following sections.

\subsection{Ward identities}
Equation \eqref{eq:oneptfns} expresses the one-point function of the energy-momentum tensor  in terms of the normalizable modes $h_{(2)ij}$. In the asymptotic analysis of the equation of motion we obtained the constraints \eqref{eq:h2constraints} for the coefficients $h_{(2)ij}$. We may now translate these equations into Ward identities involving the energy-momentum tensor. For example, the first equation in \eqref{eq:h2constraints} leads directly to the trace anomaly:
\be
\label{eq:traceanomalyb0}
\vev{T_i^i} = \mathcal A = 2 \pi \tilde R[h_{(0)}] = \frac{c}{12} \tilde R[h_{(0)}]\,.   
\ee
Reinstating the factor of $(16 \pi G_N)^{-1}$, we find $c = 3/(2G_N)$, as expected \cite{Brown:1986nw,Henningson:1998gx,Henningson:1998ey}. Furthermore, one easily finds that the last two equations in \eqref{eq:h2constraints} become the usual conservation equations:
\be
\del^i \vev{T_{ij}} = 0 
\ee
where we also used the first equation in \eqref{eq:h2constraints}.

\subsection{Callan-Symanzik equation}
In \cite{baltlogs} the holographic derivation of the Callan-Symanzik equation for scale transformation of the on-shell action was explained in complete detail. Although we will need such a detailed analysis at higher orders in $b$, at this order the analysis is familiar and essentially dates back to \cite{Henningson:1998gx,Henningson:1998ey}. Following the known prcoedures, then, we find that the Callan-Symanzik equation takes the form:
\be
\label{eq:cseqn}
\int d^2 x \Big( 2 \big(g_{(0)ij} +  h_{(0)ij}\big) \fdel{h_{(0)ij}} + 2 a_{(0)i} \fdel{a_{(0)i}} \Big) S_{\text{ren}} = - \mathcal B\,.
\ee
Here the integrated conformal anomaly is denoted $\mathcal B$. It originates from the manifest radial dependence of the counterterm action, more precisely
\be
\label{eq:CSanomalyb0}
\mathcal B = \lim_{r \to \infty} \frac{\del}{\del r} S_{\text{ct}} = \int d^2 x \Big(\qt a^k_{(0)}\square^2 a_{(0)k} - \hf \del^k a_{(0)k} \square \del^l a_{(0)l}\Big)\,,
\ee
where the partial derivative with respect to $r$ is taken with all the bulk fields fixed. On the left-hand side of \eqref{eq:cseqn} all the sources scale with their canonical dimension, so there are no beta functions at this order.

Notice that the integral of the conformal anomaly $\mathcal A$ appearing in \eqref{eq:traceanomalyb0} should in principle be equal to $\mathcal B$ as given in \eqref{eq:CSanomalyb0}. However in perturbation theory the two are manifestly different. This is firstly because the terms appearing in $\mathcal A$ are topological invariants (and we will also find total derivatives at higher orders below) and therefore integrate to zero in $\mathcal B$, and secondly because the terms appearing in $\mathcal B$ are all second order in the sources and therefore do not show up in $\mathcal A$, which in the linearized analysis is only determined to first order. In perturbation theory, then, the Callan-Symanzik equation and the trace Ward identity are both valuable and complementary sources of information. 


\subsection{Two-point functions}
The above analysis relied only on the asymptotic analysis of the equation of motion, which is always sufficient to obtain the Ward identitites and the Callan-Symanzik equation. On the other hand, to compute two-point functions we need to demand regularity inside the bulk as well. For the massive vector field the solution to the linearized equations of motion is given in terms of Bessel functions. More precisely, we find the regular solution to the equation of motion to be:
\be
a_{u} = a_{(0)u} q_u q_v K_{0}(\sqrt{2 q_u q_v} e^{-r})  + \qt \sqrt{2 q_u q_v} e^{-r} \Big(a_{(0)u} (4e^{2r} + q_u q_v) + q_u^2 a_{(0)v}\Big) K_{1}(\sqrt{2 q_u q_v} e^{-r})
\ee
and similarly with $u$ and $v$ interchanged. We can expand the Bessel functions,
\be
\begin{split}
K_{0}(z) &=\Big(- \gamma + \qt (1 - \g)z^2 + \ldots \Big) + \log(z/2) \Big( 1 + \qt z^2 + \frac{1}{64} z^4 + \ldots \Big)\\
K_{1}(z) &= \Big(\frac{1}{z} + \qt (2 \g - 1)z + \ldots \Big) + \frac{z}{2}\log(z/2)\Big( 1 + \frac{1}{8} z^2 + \frac{1}{192}z^4 + \ldots \Big)
\end{split}  
\ee
and compare with \eqref{eq:aexpansion} to find an exact match for the leading terms and to also find that
\be
a_{(4)u} = 
 \frac{1}{8}  q_u^3 q_v (\ln(q_u q_v) - \ln(2)  -1  + 2 \g) a_{(0)v} + \frac{1}{16} a_{(0)u} q_u^2 q_v^2\,. 
\ee
We may now substitute this expression the first equation in \eqref{eq:oneptfns} and take another functional derivative to obtain the two-point functions. In conventions where
\be
\op_i(x) = \int \frac{d^2 q}{(2\pi)^2} \op_i(q) e^{i q \cdot x}  \qquad \qquad  \vev{\op_i \ldots} = i \fdel{a_{(0)}^i} \vev{\ldots}\,,
\ee
with $q \cdot x = q_u u +q_v v$ and the factor of $i$ being a consequence of the Lorentzian signature (see appendix B of \cite{Skenderis:2009nt}), we find that, up to contact terms:
\be
\vev{\op_u(u,v) \op_u(0)} =  - \frac{i}{2} \int \frac{d^2 q}{(2\pi)^2}  q^3_u q_v \ln(q_u q_v - i \e) e^{i q \cdot x} =  - \frac{3}{2 \pi u^4 v^2}
\ee
and of course similarly with $u$ and $v$ interchanged. Notice how we inserted an $i \epsilon$ prescription in this correlation function. This particular $i \epsilon$ prescription corresponds to a time-ordered two-point function. We will be computing exclusively time-ordered two-point functions in this paper. The $i \epsilon$ prescription can in principle be found from a careful bulk analysis, see \cite{Skenderis:2008dh,Skenderis:2008dg} for details, the end result of which will be precisely the insertion as we have written it.\footnote{Another discussion of the real-time subtleties in Schr\"odinger spacetimes can be found in \cite{Leigh:2009eb}.} To enhance readability we have omitted the corresponding $i \epsilon$ prescription from the position-space expression, for time-ordered two-point functions it is easily reinstated by replacing $t \to t - i \e t$.

For the energy-momentum tensor the normalizable coefficients $h_{(2)ij}$ are in fact completely fixed by the Ward identities, which is a reflection of the fact that three-dimensional gravity has no propagating degrees of freedom. We find that
\be
h_{(2)uu} = \frac{q_u}{q_v}h_{(2)uv} = \frac{1}{4} ( \frac{q_u^3}{q_v} h_{(0)vv} +  q_u q_v h_{(0)uu} - 2 q_u^2 h_{(0)uv})
\ee
and therefore, using that in Lorentzian signature insertions of the energy-momentum tensor are defined via
\be
\vev{T^{ij} \ldots} = - 4 \pi i \fdel{h_{ij}} \vev{\ldots}\,,
\ee
we find that up to contact terms,
\be
\label{eq:twopttuub0}
\vev{T_{uu}(u,v) T_{uu}(0)} = - 4 \pi^2 i \int \frac{d^2 q}{(2\pi)^2} \frac{q_u^4}{q_u q_v - i \e} e^{i q \cdot x} = \frac{12 \pi} {u^4}\,, 
\ee
and again similar results hold with $u$ and $v$ interchanged. The coefficient of the two-point function is $c /2 = 3 / (4 G_N)$ if we reinstate the factor $16 \pi G_N$ again, in agreement with expectations. Notice that we rewrote $q_v^{-1}$ as $q_u/ (q_u q_v - i \epsilon)$. Such a prescription should always be understood when we write inverse powers of momenta since it is the correct regularization of the singularity at $q_v = 0$ for time-ordered correlation functions.

\section{Asymptotic behavior for finite $b$}
\label{sec:asymptsol}
As we emphasized above, our approach will be to treat the parameter $b$ as infinitesimally small and to work perturbatively in $b$. In this section we however first present the solution to the linearized equation of motion for finite $b$. Apart from certain technical complications, already discussed in \cite{us}, the main conceptual difficulty is the proper redefinition of the integration constants to what will eventually be the source and vev terms in the dual field theory. We shall demonstrate that this redefinition can be performed by taking a careful $b \to 0$ limit.

\subsection{Asymptotic solution}
For finite $b$ the linearized equations of motion are rather complicated. Exact expressions for these equations can be found in \cite{us}, where the asymptotic solution was also presented. Here we will review and slightly rewrite this asymptotic solution.

We first consider the radial component $a_r$ of the vector field. Regarding the dynamical  part of $a_r$, we find that we can combine the equations of motion to write a single fourth-order equation involving only $a_r$ of the form:
\be
\begin{split}
&4 q_u^2 q_v^2 a_r+4 e^{2 r} q_u q_v \left(b^2 q_v^2 a_r-a_r''\right)\\
& \qquad \qquad +e^{4 r} \left(-16 a_r'-4 a_r''+b^2 q_v^2 \left(\left(8+b^2 q_v^2\right) a_r -2 \left(2 a_r'+a_r''\right)\right)+4 a_r'''+a_r''''\right) = 0
\end{split}
\ee
where a prime denotes a derivative with respect to $r$. This equation can be solved in terms of Bessel functions. In addition to the solution to this equation there is also a non-dynamical part of $a_r$ which comes from a linearized diffeomorphism. The most general solution for $a_r$ is then of the form:
\be
\label{eq:arbessels}
a_{r} = \frac{i b}{q_v} \c_{(2)vv} + c_1 \mathcal I_{1}(e^{-2r} q_u q_v) + c_{-1} \mathcal I_{-1}(e^{-2r} q_u q_v) + c_9 \mathcal I_{9}(e^{-2r} q_u q_v) + c_{-9} \mathcal I_{-9}(e^{-2r} q_u q_v)
\ee
where $\c_{(2)vv}$, $c_{\pm 1}$ and the $c_{\pm 9}$ are five integration constants, which a priori can be arbitrary functions of $q_u$ and $q_v$ but not of $r$. (The notation $\c_{(2)vv}$ will be useful below.) The functions $\mathcal I_{\pm 1}(x)$ and $\mathcal I_{\pm 9}(x)$ take the form:
\be
\begin{split}
\mathcal I_{\pm 1} (x) &\equiv \sqrt{2 x} (s_1 \pm 1) I_{\pm s_1}(\sqrt{2x}) \mp I_{-1 \pm s_1}(\sqrt{2x})\\
\mathcal I_{\pm 9} (x) &\equiv \sqrt{2x} (s_9 \mp 3) I_{\pm s_9}(\sqrt{2x}) \mp I_{-1 \pm s_9}(\sqrt{2x})
\end{split}
\ee
with
\be
s_1 = \sqrt{1 + b^2 q_v^2} \qquad \qquad s_9 = \sqrt{9 + b^2 q_v^2}
\ee
and $I_\nu(z)$ the modified Bessel function of the first kind. Recalling the series expansion of the Bessel function,
\be
\label{eq:expbessel}
I_\nu(z) = 2^{-\nu} z^\nu \sum_{k=0}^\infty \frac{1}{4^k \Gamma(\n + k + 1) k!} z^{2k}  
\ee
we see that the four solutions have asymptotic behavior:
\be
\mathcal I_{\pm 1}(e^{-2r} q_u q_v) \sim e^{-(1\pm s_1)r} + \ldots \qquad \qquad \mathcal I_{\pm 9}(e^{-2r} q_u q_v) \sim e^{-(1\pm s_9)r} + \ldots 
\ee
where the dots represent infinite series which trail the leading term by powers of $\exp(-2r)$. Notice the rather unusual behavior where the power of $\exp(r)$ becomes dependent on the lightcone momentum $q_v$. This phenomenon is typical for the Schr\"odinger spacetimes and shows that the anomalous dimensions in the dual theory become dependent on the momenta.

As usual, regularity inside the bulk will eventually require us to pick specific linear combinations of these four Bessel functions. More precisely, the regularity condition will lead to two linear relations between the four coefficients $c_{\pm 1}$ and $c_{\pm 9}$. It is however important that we only impose such a regularity condition \emph{after} finishing the asymptotic analysis. This is because the holographic renormalization procedure is equivalent to finding the correct variational principle for the gravitational dynamics \cite{Papadimitriou:2005ii}. As in ordinary classical mechanics, obtaining such a variational principle requires one to treat the conjugate momenta as unrelated to the position variables. On a practical level, regarding the source and vev terms as independent becomes essential for obtaining correct results in the presence of multi-trace operator mixing \cite{baltlogs}. For these reasons we will not yet impose regularity in the interior.

Using the above solution for $a_r$ one can easily reconstruct the solution for the other component fields using the remaining equations of motion. For the metric components we find that: 
\be
\begin{split}
\label{eq:hijradexp}
h_{vv} &= e^{2r} {\color{red}\chi_{(0)vv}} + \chi_{(2)vv} \\
h_{uv} &= - \hf b^2 e^{4r} \c_{(0)vv} - b^2 e^{2r} r \c_{(2)vv} + e^{2r} {\color{red}\c_{(0)uv}} + \c_{(2)uv} \\
h_{uu} &= b^2 e^{4r} \frac{1}{q_v^2} (-q_u q_v\c_{(0)vv} + 4 \c_{(2)vv})  - 2 b^2 e^{2r} r \frac{q_u}{q_v} \c_{(2)vv} + e^{2r} {\color{red}\c_{(0)uu}} + \c_{(2)uu}\\ &\qquad 
+ c_{1} \mathcal H_{1}(e^{-2r} q_u q_v) +c_{-1} \mathcal H_{-1} (e^{-2r} q_u q_v)  
\end{split}
\ee
where we introduced five new integration constants $\c_{(0)uu}$, $\c_{(0)uv}$, $\c_{(0)vv}$, $\c_{(2)uu}$ and $\c_{(2)uv}$, whilst $\c_{(2)vv}$ and $c_{\pm 1}$ were already introduced above. The highlighting will be explained later but we warn the reader that the colored coefficients will \emph{not} be equal to the sources of the boundary theory. The functions $\mathcal H_{\pm 1}$ are slight variations of the combination of Bessel functions $\mathcal I_{\pm 1}$, namely
\be
\begin{split}
\mathcal H_{\pm 1}(x) &= \mp 8 i \sqrt{2} b\, q_u^2 q_v x^{-5/2} ( 6 + 2 b^2 q_v^2 \mp 6 s_1 + x) I_{\pm s_1-4}(\sqrt{2x})\\
& \qquad + 32 i b \, q_u^2 q_v x^{-3} \left( b^2 q_v^2 \left(s_1 \mp 6\right)+12 \left(s_1\mp 1\right)+\left(s_1 \mp 2\right) x \right) I_{\pm s_1 -3}(\sqrt{2x})\,.
\end{split}
\ee
Furthermore, the coefficients $\c_{(2)ij}$ are related to the $\c_{(0)ij}$ via:
\be
\label{eq:chi2constraints}
\begin{split}
&\c_{(2)uv} = \qt(q_v^2 \c_{(0)uu} - 2 q_u q_v \c_{(0)uv} + q_u^2 \c_{(0)vv})\\
&q_v \c_{(2)uv} - q_u \c_{(2)vv} = 0\\
&q_u \c_{(2)uv} - q_v \c_{(2)uu} = 0
\end{split}
\ee
which happens to be identical to \eqref{eq:h2constraints} in the undeformed theory.

For later reference we will give the asymptotic expansion of $h_{uu}$. Using \eqref{eq:expbessel} we find that:
\be
\begin{split}
&h_{uu} = b^2 e^{4r} \frac{1}{q_v^2} (-q_u q_v\c_{(0)vv} + 4 \c_{(2)vv})  - 2 b^2 e^{2r} r \frac{q_u}{q_v} \c_{(2)vv} + e^{2r} \c_{(0)uu}\\
&\quad+ e^{r(3 - s_1)} c_1 \left( - \frac{i 2^{\frac{7}{2} - \hf s_1} b (q_u q_v)^{(1+ s_1)/2}}{q_v s_1 (3 + b^2 q_v^2- 3 s_1) \Gamma(s_1 - 2)} - \frac{i  2^{\frac{5}{2} - \hf s_1} (s_1 -1) (q_u q_v)^{(3+s_1)/2} }{ b q_v^3 \G(1 + s_1)} e^{-2r} + \ldots \right)\\
&\quad+ e^{r(3 + s_1)} c_{-1} \left( - \frac{i 2^{\frac{7}{2} + \hf s_1} b (q_u q_v)^{(1- s_1)/2}}{q_v s_1 (3 + b^2 q_v^2+ 3 s_1) \Gamma(-s_1 - 2)} + \frac{i  2^{\frac{5}{2} + \hf s_1} (- s_1 -1) (q_u q_v)^{(3-s_1)/2} }{ b q_v^3 \G(1 - s_1)} e^{-2r} \right.\\
&\qquad \qquad \qquad \qquad \qquad \left.- \frac{i 2^{\hf + \hf s_1} (q_u q_v)^{(5 - s_1)/2}}{b q_v^3 s_1 (-3 + b^2 q_v^2) \G(-s_1 - 2)}e^{-4r} + \ldots \right)\,.
\end{split}
\ee


For $a_u$ and $a_v$ we find similar expressions as for $a_r$ which can again be written in terms of four Bessel functions. We will present only their asymptotic expansions here. These are given by: 
\be
\label{eq:auradexp}
\begin{split}
&a_u =
e^{(1 - s_1)r} \psi_{(4)v} \left(\frac{-2 (s_1 - 1)}{q_v^2} + \frac{q_u^2(-2 + s_1)}{4 (-3  +b^2 q_v^2)} e^{-4r}  + \ldots \right) \\
&+ e^{(1 + s_1)r} {\color{red}\psi_{(0)u}}\left(1 - \frac{q_u^2 (-1 + b^2 q_v^2 + s_1)}{8 b^2 ( b^2 q_v^2 - 3)} e^{-4r} + \ldots \right) \\
&+ e^{(1-s_9)r}  \psi_{(4)u} \left(1  + \frac{q_u q_v}{1 + s_9} e^{-2r} + \frac{q_u^2 q_v^2( 3 b^2 q_v^2 + 5 + 5 s_9)}{8 (11+ b^2 q_v^2+ 3 s_9)(3 + b^2 q_v^2 + s_9)} e^{-4r} + \ldots \right)\\
&-  e^{(1+ s_9)r} \psi_{(0)v} \left( \frac{2(s_9 - 3)}{q_v^2} + \frac{2 q_u(18 - 6 s_9 + b^2 q_v^2(2 -s_9)) }{q_v (8 + b^2 q_v^2) s_9} e^{-2r} + \frac{q_u^2 (16 +  3 b^2 q_v^2  - 8 s_9)}{4 (8+ b^2 q_v^2)(-2 +  s_9)} e^{-4r}+ \ldots \right) \\
& \qquad +
 e^{2r} \left(\frac{b q_u}{2 q_v} \c_{(0)vv} - \frac{2 b}{q_v^2} \c_{(2)vv} \right) + \hf b  \c_{(2)uv}
\end{split} 
\ee
and
\be
\label{eq:avradexp}
\begin{split}
a_v &= e^{-(1+ s_1)r} \psi_{(4)v} \left(1 + \frac{q_u(-1 + s_1)}{2 b^2 q_v} e^{-2r} + \ldots\right)\\
&+ e^{- (1 - s_1)r} \psi_{(0)u} \left(\frac{q_v^2}{2(1+ s_1)} - \frac{q_u q_v}{4 b^2} e^{-2r} + \ldots\right)\\
&+ e^{- (1 + s_9)r} \psi_{(4)u}\left(\frac{q_v^2(s_9 - 2)}{2 (3 + b^2 q_v^2 + s_9)} + \frac{q_u q_v^3(-2 + s_9)}{4(1+ s_9) (3 + b^2 q_v^2 + s_9)} e^{-2r} + \ldots\right)\\
&+ e^{(s_9 - 1)r} {\color{red}\psi_{(0)v}} \left(1 - \frac{q_u q_v (1 + s_9)}{2 (8 + b^2 q_v)} e^{-2r} + \ldots\right)\\
& \qquad + \hf b e^{2r} \c_{(0)vv} + \hf b \c_{(2)vv}\,.
\end{split}
\ee
Here we introduced new integration constants $\psi_{(0)i}$ and $\psi_{(4)i}$ which are a related to the $c_{\pm 1}$ and $c_{\pm 9}$ as:
\be
\label{eq:psis}
\begin{split}
\psi_{(0)u} &= -\frac{2^{\frac{1}{2} \left(1+s_1\right)} i q_v (q_u q_v)^{\frac{1}{2} \left(1-s_1\right)} }{\G(1-s_1)} c_{-1}\\
\psi_{(0)v} &= - \frac{2^{\frac{1}{2} \left(1+s_9\right)} i q_v (q_u q_v)^{\frac{1}{2} \left(1-s_9\right)}}{\G(1-s_9)} c_{-9}\\
\psi_{(4)u} &= -\frac{2^{\frac{3}{2}-\frac{1}{2} s_9} (q_u q_v)^{\frac{1}{2} (1+s_9)} \left(3 +b^2 q_v^2 +s_9\right)}{i q_v \left(-2+s_9\right) \G(1+s_9)} c_{9}\\
\psi_{(4)v} &= \frac{2^{\frac{1}{2}-\frac{1}{2} s_1} i q_v (q_u q_v)^{\frac{1}{2} \left(1+s_1\right)} }{\G(1+s_1)} c_{1}\,.
\end{split}
\ee
We will henceforth use the integration constants $\psi_{(0)i}$ and $\psi_{(4)i}$ rather than $c_{\pm 1}$ and $c_{\pm 9}$.

\subsection{Interpretation}
\label{subsec:interpret}
The solution presented in the previous subsection appears rather unstructured. In contrast to the $b=0$ analysis of section \ref{sec:linb0}, where every component of the various fields had a simple expansion in powers of $\exp(-2r)$, here we find several expansions that appear simultaneously in various components of the fields. Let us now discuss how to perform the holographic analysis in this case. 

Our first task is to identify the terms in the radial expansion that correspond to the sources of the bulk theory. This would be a trivial exercise in the usual AdS/CFT setup, where the sources are usually just the leading terms in the radial expansion of the fields, at least up to well-known subtleties for operators with $\D < d/2$ discussed in \cite{Klebanov:1999tb} and up to multi-trace deformations discussed in for example \cite{Papadimitriou:2007sj,Berkooz:2002ug,Witten:2001ua,Elitzur:2005kz,Mueck:2002gm,Sever:2002fk,Akhmedov:2002gq,baltlogs}. In our case inspection of the solution naturally leads to the idea that the $\psi_{(0)i}$ and $\c_{(0)ij}$ could be sources in a dual field theory. These are the coefficients marked in red in the above formulas. Indeed, the $\psi_{(0)i}$ are leading terms in an infinite expansion in powers of $\exp(-2r)$ and upon sending $b \to 0$ the power of $\exp(r)$ accompanying the coefficients $\psi_{(0)u}$ and $\psi_{(0)v}$ becomes precisely $\exp(2r)$ in $a_u$ and $a_v$, respectively, since in this limit $s_1 \to 1$ and $s_9 \to 3$. The $\psi_{(0)i}$ therefore `land' precisely at the order of the source terms in the $b=0$ expansion of the previous section. Similarly, the $\c_{(0)ij}$ appear precisely at the same order, $\exp(2r)$, as the sources in the $b=0$ case. They might therefore correspond to the energy-momentum tensor in the dual theory. All the other terms in the radial expansion of the fields would then merely be `induced' terms which are sometimes more leading than the source terms \cite{balt}.  

There is however a problem with the reasoning of the previous paragraph, because upon further inspection the $b \to 0$ limit is not as straightforward as our previous discussion seems to suggest. For example, the first subleading term in the expansion multiplying $\psi_{(0)v}$ in the expansion \eqref{eq:auradexp} for $a_u$ also becomes of order $\exp(2r)$ as $b \to 0$ and will therefore mix with $\psi_{(0)u}$ to become the actual source term $a_{(0)u}$. A similar mixing can be seen to occur for many other coefficients in the radial expansion of the fields. The terms highlighted in red in the above formulas therefore do \emph{not} precisely reduce to the source terms as $b \to 0$.

This observation corroborates nicely with the results of \cite{balt, baltlogs}. As we mentioned above we view the appearance of $b$ as an irrelevant perturbation. In \cite{balt} it was demonstrated that \emph{to any finite order in the irrelevant deformation parameter, the source is always equal to the term at order $\exp(-\D_- r)$ in the radial expansion of the fields}. Here $\D_-$ is the appropriate scaling dimension for the dual source \emph{without} the deformation, which in our case happens to be equal to $-2$ for both the metric and the vector field. This simple prescription leads one directly to the correct identification of the sources, as we will now proceed to demonstrate. 

\subsection{Identification of the sources}
The result of \cite{balt} is valid perturbatively in the irrelevant deformation parameter, \ie in $b$. This in particular implies that the $b$-dependent powers of $\exp(-r)$, so terms like $\exp(\pm s_9 r)$, should be viewed as infinite series expansions in $b$ before the analysis of \cite{balt} can be applied. In order to identify the proper sources we should therefore write such terms as:
\be
\exp(\pm s_9 r) = \exp\big(\pm \sqrt{9 + b^2 q_v^2} \, r\big) \to \exp(\pm 3 r) \left( 1 \pm \left(\frac{1}{6} b^2 q_v^2 - \frac{1}{216} b^4 q_v^4\right) r + \frac{1}{72} b^4 q_v^4 r^2 + \ldots \right)  
\ee   
so as a term of order $\exp(\pm 3 r)$ plus an infinite series of logarithmic terms.\footnote{We call terms involving powers of our radial coordinate $r$ `logarithmic' because they are powers of logarithms of the conventional Fefferman-Graham radial coordinate $\rho = \exp(-r)$ and also because they correspond to beta functions and conformal anomalies in the dual field theory.} As explained in \cite{baltlogs}, the presence of these logarithmic terms does not modify the identification of the source as the term multiplying simply the $\exp(- \D_- r)$ coefficient. For the identification of the sources it is then sufficient to replace $\exp(\pm s_9 r)$ with $\exp(\pm 3r)$. Of course the same analysis can be applied to the other $b$-dependent powers, so:
\be
\exp(\pm s_1 r) = \exp\big(\pm \sqrt{1 + b^2 q_v^2} \, r\big)  \to \exp(\pm r) \left( 1 \pm \left( \hf b^2 q_v^2 - \frac{1}{8} b^4 q_v^4\right) r + \frac{1}{8} b^4 q_v^4 r^2 + \ldots \right)\,.
\ee

Taking into account such an interpretation of the fractional powers of $\exp(r)$ allows us to directly read off the correct source terms in the linearized solutions. For example, from \eqref{eq:auradexp} and \eqref{eq:avradexp} we directly find that the terms of order $\exp(2r)$ in the radial expansion of the $a_i$ are:
\be
\label{eq:redefna0}
\begin{split}
a_{(0)u} &\equiv \psi_{(0)u} - \frac{2 q_u(18 - 6 s_9 + b^2 q_v^2(2 -s_9)) }{q_v (8 + b^2 q_v^2) s_9} \psi_{(0)v} + \frac{b q_u}{2 q_v} \c_{(0)vv} - \frac{2b}{q_v^2} \c_{(2)vv}\\
a_{(0)v} &\equiv \psi_{(0)v} + \hf b \c_{(0)vv}\,.
\end{split}
\ee
It will be useful to also define the corresponding `normalizable' terms, which by definition sit at order $\exp(-2r)$ in the radial expansion of the fields:
\be
\label{eq:redefna4}
\begin{split}
a_{(4)u} &\equiv \psi_{(4)u} + \frac{q_u^2 q_v^2}{8(1 - b^2 q_v^2 + s_1)} \psi_{(0)u} + \frac{q_u^3 q_v (2 b^2 q_v^2 - 5 s_9) (1 + s_9)}{12 (5 + b^2 q_v^2)(8 + b^2 q_v^2) (-3 + s_9)} \psi_{(0)v}\\
a_{(4)v} &\equiv \psi_{(4)v} + \frac{q_u^2 q_v^2}{88 + 8 b^2 q_v^2 - 24 s_9} \psi_{(0)v} - \frac{q_u q_v}{4b^2} \psi_{(0)u}\,.
\end{split}
\ee
For the metric we find that:
\be
\label{eq:redefnh0}
\begin{split}
h_{(0)uu} &\equiv \c_{(0)uu} - \frac{8 b}{q_v^2} \psi_{(4)v} + \frac{2 q_u}{b q_v} \psi_{(0)u}\\
h_{(0)uv} &\equiv \c_{(0)uv}\\
h_{(0)vv} &\equiv \c_{(0)vv}
\end{split}
\ee
are the correct sources in the dual theory, whereas the `normalizable' terms are given by the terms at order one:
\be
\label{eq:redefnh2}
\begin{split}
h_{(2)uu} &\equiv \c_{(2)uu} + \frac{q_u^2}{2b ( 2 - s_1)} \psi_{(0)u} - \frac{4 q_u (s_1 - 1)}{b q_v^3} \psi_{(4)v}\\
h_{(2)uv} &\equiv \c_{(2)uv}\\
h_{(2)vv} &\equiv \c_{(2)vv}\,.
\end{split}
\ee

The above transformation between the correct field theory sources, so $h_{(0)ij}$ and $a_{(0)i}$, and the natural integration constants, so $\chi_{(0)ij}$ and $\psi_{(0)i}$, is quite remarkable. Indeed, the modes parametrized by the $\chi_{ij}$ represent a linearized diffeomorphism whereas the modes parametrized by the $\psi_{i}$ represent a propagating fluctuation. They are thus qualitatively very different and one may naturally try to associate them with different operators in the dual theory. We however now observe that the correct field theory sources $h_{(0)ij}$ and $a_{(0)i}$ are \emph{not} in one-to-one correspondence with the two sets of solutions parametrized by the $\chi_{ij}$ and the $\psi_{i}$. Rather, each of the field theory sources switches on a combination of these two solutions. This is in fact reminsicent of the analysis of TMG at the chiral point presented in \cite{Skenderis:2009nt}, where a similar mixing between the propagating solution (\ie the Bessel function) and the linearized diffeomorphism occurred. 

We would like to emphasize that the redefinitions \eqref{eq:redefna0} and \eqref{eq:redefnh0} are very non-local because of the factor of $q_v^{-1}$ and $q_v^{-2}$ appearing in the transformation formulae. Notice that, despite appearances, we are using an $i \e$ prescription such that these factors are in fact non-local in the $u$ direction, see for example the two-point function \eqref{eq:twopttuub0}. Any analysis where the $\chi_{(0)ij}$ and $\psi_{(0)i}$ are treated as the field theory sources will therefore hide a non-local convolution of the correct field theory correlation functions and may easily lead to erroneous conclusions.\footnote{Although the field redefinition from the $h_{(0)ij}$ and $a_{(0)i}$ to the $\chi_{(0)ij}$ and $\psi_{(0)i}$ is non-local, there do exist many other field redefinitions which are local and therefore in principle allowed. At large $N$, the full space of such local field redefinitions also includes local double-trace deformations and Legendre transforms akin to \cite{Klebanov:1999tb}. As explained further in the remaining sections, rather than exhaustively exploring all possible redefinitions we have instead opted to fix this freedom in a rather natural way.}    

\section{Analysis to first order in $b$}
\label{sec:linb1}
Altough we have just presented the solution to the linearized equation of motion for finite values of $b$, we cannot directly use this solution to compute correlation functions since we are unable to perform the holographic renormalization procedure for finite $b$. We announced in the introduction that our strategy will be to work perturbatively in $b$ instead. In this section we will make the first step by performing the holographic analysis to first order in $b$. The structure and methods of this section are exactly the same as that of \cite{balt,baltlogs} so the presentation will be rather brief. We will however discuss the field theory results in a little more detail.
   
\subsection{Linearized asymptotic solution}
From the results of the previous section we may easily find the order $b$ correction to the linearized solution of the equations of motion. For the metric perturbation $h_{ij}$ it takes the form:
\be
\begin{split}
h\as{1}{uu} &= - 2 b a_{(0)u} e^{4r} + 2 b r q_u q_v a_{(0)u} e^{2r} + \hf b r q_u^2 q_v^2 a_{u(0)} + h\as{1}{(2)uu}  e^{-2r} + \ldots \\
h\as{1}{uv} &= h\as{1}{(2)uv} e^{-2r} \\
h\as{1}{vv} &= h\as{1}{(2)vv} e^{-2r}
\end{split}
\ee
where we added the curly bracket $\{1\}$ to indicate that this is the first-order correction. Notice that there are no (non-logarithmic) terms at order $\exp(2r)$. This corresponds to the fact that we set any order $b$ corrections to the source to zero, in agreement with the analysis of \cite{balt}. However we did allow for a correction to the normalizable terms, which we indicated with a term $h\as{1}{(2)ij}$. In general these corrections would be unknown but in three-dimensional gravity the equations of motion furthermore dictate that:
\be
\label{eq:wardidsb}
\begin{split}
&h\as{1}{(2)uv} = - \frac{1}{8} b q_u^2 q_v^2 a_{(0)v} + 2 b a_{\{0\}(4)v}\\
&q_v h\as{1}{(2)uv} - q_u h\as{1}{(2)vv} = 0\\
&q_u h\as{1}{(2)uv} - q_v h\as{1}{(2)uu} = 0
\end{split}
\ee
which will become the first-order corrections to the Ward identities of the dual field theory. The appearance of the vector field components in $h\as{1}{(2)uv}$ already indicates that the Ward identities will change at this order.

For the vector field the corrections take the form:
\be
\begin{split}
a_{\{1\}u} &= b \frac{1}{8} q_u^3 q_v h_{(0)vv} r e^{-2r} + a_{\{1\}(4)u} e^{-2r} + \ldots \\
a_{\{1\}v} &= b h\as{0}{(2)vv} - b r (\hf q_u q_v h\as{0}{(2)vv} - \frac{1}{8} q_u^2 q_v^2 h_{(0)vv}) e^{-2r} + a_{\{1\}(4)v} e^{-2r} + \ldots
\end{split}
\ee
The term $b \exp(2r)$ in $A_u$, from which all the corrections we wrote above originate, is in our conventions a change in the background vector field and therefore does not appear in the above expansion.

\subsection{Holographic renormalization}
\label{subsec:hrb1}
The computation of the counterterm action is a straightforward implementation of the general methods discussed in \cite{balt,baltlogs}. Using Mathematica we found that the counterterm action to first order in $b$ is given by:
\be
\label{eq:s1ct}
S_{\{1\}ct} = \frac{1}{2} \int d^2 x \sqrt{-\g} b^i \Big( h_{ij} \del^j \del^k a_k - r h_{ij} \square_\g \del^j \del^k a_k \Big)\,.
\ee
Here we replaced the parameter $b$ with a one-form field $b_i dx^i = b \exp(2r) du$ which allowed us to write the counterterm action in a covariant form. As usual, indices are raised and lowered and covariant derivatives are taken using the background metric $\g_{ij}$.

\subsubsection{Scheme dependence}
The counterterm action \eqref{eq:s1ct} is not unique, since we may freely add certain scheme-dependent finite counterterms as well. For example, a term of the form:
\be
\int d^2 x \sqrt{-\g} b^i h_{ij} \square_\g \del^j \del^k a_k
\ee
is finite as $r \to \infty$ and can therefore in principle be freely added to \eqref{eq:s1ct}. Although doing so will modify the correlation functions at this order in $b$ by nothing more than contact terms, at higher orders in $b$ the addition of such terms will affect the form of the counterterm action and also the Ward identities. In order to make our results meaningful at every order in $b$ we therefore have to properly identify the scheme dependence. In the next few paragraphs we explain our approach.

First of all, at every order certain finite counterterms are necessary to maintain general covariance. Since the underlying problem is generally covariant, our counterterms are in principle the expansion of covariant counterterms. For example, a volume counterterm leads to a specific expansion:
\be
\sqrt{\det{\g}} \to \sqrt{\det{\g}} \Big(1 + \hf \g^{ij} h_{ij} + \qt (\g^{ij} h_{ij})^2 - \qt \g^{ij} h_{jk} \g^{kl} h_{li} + \ldots\Big)
\ee
as $\g_{ij} \to \g_{ij} + h_{ij}$. Even though every single term on the right-hand side appears covariant, it is only the specific linear combination given above that sums to a completely covariant form.\footnote{Such relations between counterterms are not exceptional and also occur in e.g. the non-linear $O(N)$ model.} To find the relations between the counterterms such that general covariance is retained, we need to verify that we are not accidentally breaking diffeomorphism invariance. This can be easily verified by demanding that the energy-momentum tensor of the boundary theory is exactly conserved. As we will explain in the next subsection, the correct energy-momentum tensor for the deformed boundary theory is the tensor $J_{ij}$ defined in equation \eqref{eq:jij}. Demanding then that
\be
\label{eq:delijij}
\del^i J_{ij} = 0
\ee  
to any order in $b$ fixes a large amount of scheme dependence in the counterterms. Notice that $J_{ij}$ already takes into account the transformation properties of the null vector $b$ under diffeomorphisms -- indeed, diffeomorphisms in the deformed theory can of course only be unbroken if we allow ourselves to transform $b$ as well.

After having imposed general covariance we are left with a finite number of finite counterterms at every order in $b$. Together with perturbative field redefinitions these reflect the true scheme dependence that arises in any covariant renormalization method. In our computations we have identified all those counterterms and we added them with arbitrary coefficients. This allowed us to fully understand the scheme dependence at every order in $b$. We will indicate any relevant scheme dependence in our results in what follows.

\subsection{One-point functions}
From the first variation of the renormalized on-shell action we obtain the following one-point functions:
\be
\label{eq:oneptfirstb}
\begin{split}
\vev{\op_u} &= -\frac{1}{\sqrt{-g_{(0)}}} \fdel[S]{a_{(0)v}} = - 4 (a_{\{0\}(4)u} + a_{\{1\}(4)u}) + \text{(contact terms)} \\
\vev{\op_v} &= -\frac{1}{\sqrt{-g_{(0)}}} \fdel[S]{a_{(0)u}}   = - 4 (a_{\{0\}(4)v} + a_{\{1\}(4)v}) + \text{(c.t.)} \\
\vev{T_{uu}} &= \frac{4\pi}{\sqrt{-g_{(0)}}} \fdel[S]{h_{(0)vv}} = 4 \pi \Big(h_{\{0\}(2)uu} + b h_{\{1\}(2)uu} - 2 b a_{\{0\}(4)u} + \text{(c.t.)} \Big)\\
\vev{T_{uv}} &= \frac{4\pi}{\sqrt{-g_{(0)}}} \fdel[S]{h_{(0)uv}} = -4 \pi \left( h_{\{0\}(2)uv} +  h_{\{1\}(2)uv} \right)\\
\vev{T_{vv}} &= \frac{4\pi}{\sqrt{-g_{(0)}}} \fdel[S]{h_{(0)uu}} = 4 \pi \left(  h_{\{0\}(2)vv} +  h_{\{1\}(2)vv} \right)\,. 
\end{split}
\ee
Let us briefly explain the structure of these results. First of all, to avoid clutter we have chosen not to write the various contact terms since they are largely scheme-dependent. (We will reinstate the relevant contact terms in the Ward identities below.) Apart from the contact terms, then, the results at order $b^0$ are precisely those of equation \eqref{eq:oneptfns} written out in components. At order $b$ we of course see that the coefficients of the normalizable modes $h_{\{0\}(2)ij}$ and $a_{\{0\}(2)i}$ get corrected with the $h_{\{1\}(2)ij}$ and the $a_{\{1\}(4)i}$, as expected from the general arguments of \cite{balt}. We also find a new non-contact term appearing in the one-point function of $T_{uu}$. This term is indicitative of some new structure in the two-point functions, in particular the two-point function $\vev{T_{uu} \op_u}$ will be non-zero at this order. Apart from the Ward identities discussed below, this non-contact term is the main nontrivial result of a careful holographic renormalization procedure at this order. We are not aware of any other (possibly faster) way of obtaining it.
 
 \subsection{Ward identities and Callan-Symanzik equation}
Equations \eqref{eq:wardidsb} plus their order $b^0$ counterparts \eqref{eq:h2constraints} can now be used to find the Ward identities in the deformed theory. We first of all obtain that:
\be
\label{eq:diffwardidb1}
\begin{split}
0 &= q_u \vev{T_{vv}} + q_v \vev{T_{uv}} \\
0 &= q_v \vev{T_{uu} - 2 \pi b \op_u} + q_u \vev{T_{uv} - 2\pi b \op_v}\,. 
\end{split}
\ee
In agreement with general expectations discussed in \cite{us} and in the appendix to this paper, we see that the energy-momentum tensor $T_{ij}$ is no longer conserved. Instead, the correct energy-momentum tensor is the non-symmetric combination
\be
\label{eq:jij}
J_{ij} \equiv T_{ij} - 2 \pi \op_i b_j 
\ee 
which satisfies $\del^i J_{ij} = 0$ and therefore gives the expected conserved currents associated to translation invariance.

The deformation parameter $b$ also breaks the scale invariance of the dual theory. Indeed, the classical beta function for $\op_i$ leads to a modified trace Ward identity for the energy-momentum tensor,
\be
\vev{T_i^i}  -  4 \pi b \vev{\op_v} = \mathcal A =  2 \pi \tilde R[h_{(0)}]   - c_1 b q_u^2 q_v^2 a_{(0)v} + c_2 b q_u q_v^3 a_{(0)u}\,,
\ee
with scheme-dependent coefficients $c_1$ and $c_2$. In the appendix we discuss how the left-hand side is indeed the expected trace Ward identity for a deformation with a vector operator with scaling dimension $\D = 3$ in $d=2$.

The Callan-Symanzik equation takes the form:
\be
\label{eq:cseqnfirstorder}
\int d^2 x \Big( 2 (g_{(0)ij} + h_{(0)ij}) \fdel{h_{(0)ij}} + 2 (b_i + a_{(0)i}) \fdel{a_{(0)i}} \Big) S_\text{ren} + \mathcal B = 0
\ee
with the integrated conformal anomaly
\be
\label{eq:CSanomaly}
\mathcal B = \hf \int d^2 q \Big( (a_{(0)v} q_u^3 q_v a_{(0)v} + a_{(0)u} q_u q_v^3 a_{(0)u}) - b h_{(0)vv}q_u^2 q_v (q_u a_{(0)v}  - q_v a_{(0)u})  + b \tilde R[h_{(0)}] q_v^2 a_{(0)u}\Big)\,.
\ee
The new terms in the conformal anomaly are scheme-independent. From equation \eqref{eq:cseqnfirstorder} we read off that the sources still scale with the same scaling dimension as for $b = 0$ so there are no beta functions at this order.

We will present a more extended discussion of the Ward identities and the Callan-Symanzik equation for finite $b$ in the next section. 

\subsection{Two-point functions}
As before, the two-point functions can be found by expressing the normalizable modes $a_{(4)i}$ and $h_{(2)ij}$ in terms of the sources $a_{(0)i}$ and $h_{(0)ij}$, substituting in \eqref{eq:oneptfirstb} and taking another derivative with respect to the sources. Using the fact that we can solve the linearized equations of motion completely for finite $b$ (see section \ref{sec:asymptsol} for the asymptotic expansion and section \ref{sec:linb4} below for the regularity condition) we find the following first-order expressions for the normalizable modes:
\be
\begin{split}
a_{\{0\}(4)u} + a_{\{1\}(4)u} &= \frac{1}{8}q_u^3 q_v  \big(\ln(q_u q_v) - \ln(2)  -1  + 2 \g\big)\big(a_{(0)v} - \hf b h_{(0)vv}\big) \\ &\qquad + \frac{1}{16} q_u^2 q_v^2 a_{(0)u}  + \frac{1}{32} b q_u q_v^3 h_{(0)uu} - \frac{1}{16} b q_u^2 q_v^2 h_{(0)uv}\\
a_{\{0\}(4)v} + a_{\{1\}(4)v} &= \frac{1}{8}q_u q_v^3  \big(\ln(q_u q_v) - \ln(2)  -1  + 2 \g\big)\big(a_{(0)u} - b h_{(0)uv} + \frac{1}{2} b \frac{q_v}{q_u} h_{(0)uu} \big)\\ &\qquad + \frac{1}{16} q_u^2 q_v^2 a_{(0)v}  - \frac{1}{32} b q_u^2 q_v^2 h_{(0)vv}\,. 
\end{split} 
\ee
The $h_{\{1\}(2)ij}$ are again completely determined in terms of the sources by the constraints \eqref{eq:wardidsb}. Altogether this leads to the following new non-vanishing two-point functions:
\be
\begin{split}
\vev{T_{uu} \op_u} &=  - i  b \pi \int \frac{d^2 q}{(2\pi)^2} q_u^3 q_v \log(q_u q_v - i \e) e^{i q \cdot x} = - \frac{3 b}{u^4 v^2} \\
\vev{T_{vv} \op_v} &= i b \pi \int \frac{d^2 q}{(2\pi)^2} q_v^4 \log(q_u q_v - i \e) e^{i q \cdot x} =  \frac{12 b}{u v^5} \\
\vev{T_{uv} \op_v} &= - i b \pi\int \frac{d^2 q}{(2\pi)^2} q_u q_v^3 \log(q_u q_v - i \e) e^{i q \cdot x} = - \frac{3b}{u^2 v^4}\,,
\end{split}
\ee
whereas the other two-point functions are unchanged and we again ignored various contact terms. In terms of the currents $J_{ij}$ we find the following complete set of non-zero two-point functions:
\begin{align}
\vev{J_{uu} J_{uu}} &= \frac{12 \pi}{u^4} && \vev{\op_u \op_u} = \frac{-3}{2 \pi u^4 v^2}\nonumber\\
\label{eq:twoptb1}
\vev{J_{vv} J_{vv}} &= \frac{12 \pi}{v^4} && \vev{\op_v \op_v} = \frac{-3}{2 \pi u^2 v^4}\\
\vev{J_{uv} \op_v} &= - \frac{3b}{u^2 v^4} && \vev{J_{vv} \op_v} = \frac{12 b}{u v^5}\,.\nonumber
\end{align}
The only new correlation functions are those on the last line. The first of these indicates that, although $\vev{J_{uv} J_{uv}}$ vanishes up to a contact term, the descendant operator $\del_v J_{uv}$ is not a trivial operator. In the analysis of \cite{Hofman:2011zj} it was claimed that in a two-dimensional theory with non-relativistic (or `chiral') scale invariance, at least under certain assumptions, $\del_v J_{uv}$ had to be a trivial operator and the conservation equation $\del^i J_{iv}$ would then dictate that $\del_u J_{vv}$ vanished as well. This would lead to the existence of an antiholomorphic current $J_{vv}$ which together with $J_{uu}$ would lead to two infinite-dimensional symmetry algebras. The last line of \eqref{eq:twoptb1} however explicitly indicates that this scenario is not realized in the deformed theory already to first order in $b$.\footnote{For finite $b$ we will see below that the theory becomes non-local and the spectrum of the non-relativistic dilation operator becomes continuous. This explicitly violates the assumptions of \cite{Hofman:2011zj} so the results of that paper will then automatically no longer apply. See also \cite{Nakayama:2011fe} for a related discussion.}

\section{Analysis for finite $b$}
\label{sec:linb4}
The next step in the analysis is to increase the order in $b$ in which we are working. Using Mathematica we have worked out the full holographic analsyis up to and including order $b^4$. Our reason for working to this order is that all possible contractions of $b_i$ with at most two out of the fields $h_{ij}$ and $a_{i}$ are exhausted at order $b^4$, with the final structure given by:
\be
\int d^2 x\, \sqrt{\g} \, b^i b^j b^k b^l h_{ij} h_{kl}
\ee
and what remains at higher orders are infinite series expansions in the scalar operator $b^2 q_v^2$. (The above counterterm is indeed necessary at order $b^4$, its coefficient happens to be $-1/12$.) 

The analysis at order $b^4$ leads to several interesting observations. For example, up to order $b^4$ we find from equation \eqref{eq:1ptmetricfull} below that there is no explicit term of the form $b_u a_{(4)v}$ in the one-point function of $T_{uv}$, even though such a term would be allowed by conformal and Lorentz symmetry. Although our analysis does not exclude the appearance of such a term at order $b^5$, for example with an extra factor $b^4 q_v^4$, its absence at the lowest two orders leads us to believe that this term will be absent to all orders. Using this and similar observations, we will in this section not only present the results of the perturbative analysis but also conjecture a consistent extrapolation of our results that should be valid to all orders in $b$. The equations in this section represent the results of this extrapolation, whilst the truncation of the results to order $b^4$ is based on explicit computations.

The discussion below in principle follows the pattern of sections \ref{sec:linb0} and \ref{sec:linb1}. We have however already extensively discussed the asymptotic solution to the linearized equations of motion in section \ref{sec:asymptsol} and we will therefore proceed with a discussion of the holographic renormalization procedure.

In the previous section we indicated the order $b$ correction to for example $h_{(2)ij}$ by $h_{\{1\}(2)ij}$ and correspondingly introduced $h_{\{0\}(2)ij}$ to indicate the order $b^0$ result. In this section we will drop the curly brackets and simply write $h_{(2)ij}$ to denote the all-order result, so:
\be
h_{(2)ij} = h_{\{0\}(2)ij} + h_{\{1\}(2)ij} + h_{\{2\}(2)ij} + \ldots
\ee
and we employ a similar change of notation for other coefficients.

\subsection{Holographic renormalization}
As we mentioned above, the procedure for obtaining the counterterm action in the presence of irrelevant deformations was explained in our earlier work \cite{balt,baltlogs}. With the aid of Mathematica we obtained the full counterterm action to order $b^4$, using the methods explained in subsection \ref{subsec:hrb1} to ensure covariance of the counterterm action. Furthermore, at each order we have determined all the possible finite local counterterms in order to be able to parametrize all the scheme dependence in our results.

The full form of the counterterm action is rather complicated and not too illuminating so we will not present it here. Let us however mention that, starting at order $b^2$, we also need double-trace counterterms. The full double-trace part of the counterterm action to order $b^4$ takes the form:
\be
\begin{split}
S_{\text{ct,dt}} &= \int d^2 x \sqrt{-\g} \, (r - c_0) b^i b^j \s_{ik} \s^k_j \\ 
&\qquad  - \int d^2 x \sqrt{-\g} \,(r - c_1) b^i b^j b^k \pi_i \s_{jk} \\
& \qquad + \int d^2 x \sqrt{-\g}\,
 \big(3 r - \frac{2}{3} - c_0 - 2 c_1\big) (b^i b^j \s_{ij})^2\\
& \qquad  - \int d^2 x \sqrt{-\g} \, b^i b^j b^k b^l  \left( \big(\hf r^2 - r \big(\frac{5}{12} +  c_1\big) + \frac{1}{12} c_2\big)  \s_{im}\del_j \del_k \s^m_l + c_3 \s_{ij} \square_\g \s_{kl} \right) 
\end{split}
\ee 
where the $c_i$ are scheme-dependent coefficients and $\s_{ij}$ and $\pi_{i}$ are the renormalized momenta conjugate to $h_{ij}$ and $a_{i}$, respectively. These are defined as \cite{baltlogs}:
\be
\s^{ij} = \fdel{h_{ij}}\big(S_{\text{bare}} + S_\text{ct,st}) \qquad \qquad \pi^{i} = \fdel{a_{i}}\big(S_{\text{bare}} + S_\text{ct,st})
\ee
with $S_\text{ct,st}$ the single-trace piece of the counterterm action. Notice that most of these counterterms are logarithmic in nature, \ie they involve explicit powers of $r$, leading to operator mixing between single- and double-trace operators in the Callan-Symanzik equation below. There are no higher-trace counterterms since the background conjugate momenta vanish identically and we only expand the action up to second order in the fluctuations. 

As explained in \cite{baltlogs}, logarithmic double-trace counterterms generally force one to perform field redefinitions in order to retain a correct variational principle. Indeed, for example at order $b^2$ we find terms in the variation of the on-shell action of the form: 
\be
\begin{split}
\delta S_{\text{ren}} &= \int d^2 x \Big(\ldots + h_{(2)uu} \d h_{(0)vv} + h_{(2)vv} \d h_{(0)uu} - 2 h_{(2)uv} \d h_{(0)uv} + \ldots\\ 
&\qquad \ldots - b^2 (1 - 2 c_0) h_{(2)uv} \d h_{(2)vv} + \ldots \Big)\,.
\end{split}
\ee
From the order $b^0$ metric variations on the first line we find that $h_{(2)vv}$ is a conjugate momentum so its variation should not appear in an action with an appropriately defined variational principle. In order to repair the variational principle at order $b^2$ we may either choose the scheme where $c_0 = 1/2$ or alternatively we redefine the sources to be:
\be
h_{(0)uv} \to h_{(0)uv} - \hf b^2 (1 - 2 c_0) h_{(2)vv} 
\ee  
so the variation $\d h_{(2)vv}$ cancels. The choice between the different prescriptions amounts to adding a finite double-trace counterterm to the theory, see \cite{baltlogs} for details.

In this paper we will opt for the simplest solution and set $c_0 = 1/2$ so no redefinition of the source terms is necessary. At higher orders we can continue this procedure and fix the finite terms in the multi-trace counterterm action precisely such that no field redefinitions are needed. This leads to the simple prescription that the sources are \emph{always} the terms of order $\exp(2r)$ in both $a_i$ and in $h_{ij}$.

\subsection{One-point functions}
Let us now present the one-point functions, defined as the first variation of the renormalized on-shell action. We find the following one-point functions for the vector operator:
\be
\label{eq:1ptvectorfull}
\begin{split}
\vev{\op_u} &= - \frac{1}{\sqrt{-g_{(0)}}} \fdel[S]{a_{(0)v}} = - 4 \big[ 1  + \frac{1}{36} b^2 q_v^2 - \frac{1}{648} b^4 q_v^4 + \ldots \big] a_{(4)u} \\&\qquad \qquad \qquad + \frac{5}{3} b^2 q_u^2 q_v \big[1+ \frac{2}{9} b^2 q_v^2 + \ldots \big] a_{(4)v} + \text{(contact terms)} \\
\vev{\op_v} &= - \frac{1}{\sqrt{-g_{(0)}}} \fdel[S]{a_{(0)u}} = - 4 \big[1 + \qt b^2 q_v^2- \frac{1}{8} b^4 q_v^4 + \ldots\big] a_{(4)v} + \text{(c.t.)}\,.
\end{split}
\ee
The contact terms here are scheme-dependent and we have omitted them. Although we have computed the on-shell action only to order $b^4$, we expect that at higher orders in $b$ no completely new terms will appear in the one-point functions. On the other hand, it seems likely that the expansions in $b^2 q_v^2$ that we enclosed in square brackets in equation \eqref{eq:1ptvectorfull} will continue indefinitely to form a non-trivial function of $b^2 q_v^2$. We notice that this function can in fact be removed in perturbation theory by a redefinition of the sources
\be
a_{(0)i}\to \big[ 1 + \# b^2 q_v^2 + \# b^4 q_v^4 + \ldots \Big] a_{(0)i} 
\ee
for some appropriate numerical coefficients we denoted with a $\#$ here. Such redefinitions are local to any finite order in $b$ and are therefore an allowed part of the usual scheme dependence in quantum field theory, complementary to the addition of finite local counterterms in order to shift the contact terms in the correlators.

For the one-point functions of the energy-momentum tensor we find that:
\be
\label{eq:1ptmetricfull}
\begin{split}
\vev{T_{uu}} &= \frac{4\pi}{\sqrt{g_{(0)}}} \fdel[S]{h_{(0)vv}} = 4 \pi \Big(h_{(2)uu}  - 2 b \big[ 1 + \frac{1}{9} b^2 q_v^2 + \ldots \big] a_{(4)u} - \frac{7}{6} b^3 \big[ 1 + \ldots \big] q_u q_v  a_{(4)v} + \text{(c.t.)} \Big)\\
\vev{T_{uv}} &= \frac{4\pi}{\sqrt{g_{(0)}}} \fdel[S]{h_{(0)uv}} = -4 \pi \left( \big[ 1 + \qt b^2 q_v^2 - \frac{1}{8} b^4 q_v^4 + \ldots \big]h_{(2)uv} + \text{(c.t.)} \right)\\
\vev{T_{vv}} &= \frac{4\pi}{\sqrt{g_{(0)}}} \fdel[S]{h_{(0)uu}} = 4 \pi \left( \big[ 1 + \qt b^2 q_v^2 - \frac{1}{8} b^4 q_v^4 + \ldots \big]h_{(2)vv} + \text{(c.t.)} \right)\,,
\end{split} 
\ee
where we again omitted the contact terms and wrote in square brackets the possibly infinite expansions in $b^2 q_v^2$. Notice that we did not find any $b^2 q_v^2$ terms multiplying $h_{(2)uu}$ in $\vev{T_{uu}}$ up to and including order $b^4$. Furthermore, the expansions in $b^2 q_v^2$ multiplying $h_{(2)uv}$ and $h_{(2)vv}$ in $\vev{T_{uv}}$ and $\vev{T_{vv}}$ are identical at least to order $b^4 q_v^4$. We will again assume that this is no accident and that this effect persists to all orders, even though we cannot currently offer a detailed explanation.

\subsection{Ward identities}
The equations \eqref{eq:chi2constraints} combined with the redefinitions \eqref{eq:redefna0}-\eqref{eq:redefnh2} result in constraints for the $h_{(2)ij}$. Using then \eqref{eq:1ptmetricfull} and \eqref{eq:1ptvectorfull} we can rewrite these constraints in terms of the one-point functions of the dual operators. This leads to:
\be
\begin{split}
0 &= q_u \vev{T_{vv}} + q_v \vev{T_{uv}}\\
0 &= q_v \vev{T_{uu} - 2 \pi b \op_u} + q_u \vev{T_{uv} - 2\pi b \op_v} \,.
\end{split}
\ee
The correct energy-momentum tensor is therefore again $J_{ij}$ which we recall was defined in \eqref{eq:jij} to be equal to $J_{ij} = T_{ij} - 2 \pi \op_i b_j$. This tensor is the conserved energy-momentum tensor,
\be
\del^i J_{ij} = 0\,,
\ee 
which we now claim will hold to all orders in $b$. (As discussed in section \ref{sec:linb1}, this conservation equation is essentially a consequence of our covariant renormalization scheme.)

The trace Ward identity again accurately reflects the breaking of scale invariance. With respect to the analysis at order $b^1$ we find that it is augmented only by contact terms: 
\be
\label{eq:tracewardid}
\begin{split}
T_i^i  -  4 \pi b \op_v = \mathcal A &=  2 \pi \big[ 1 + \# b^2 q_v^2 + \# b^4 q_v^4 + \ldots \big] \tilde R[h_{(0)}] \\
& \qquad  + b \big[ \# +  \# b^2 q_v^2  + \ldots \big] q_u^2 q_v^2 a_{(0)v} \\
& \qquad+  b \big[ \# + \# b^2 q_v^2 + \ldots \big]q_u q_v^3 a_{(0)u}
\\
& \qquad+ b^2  \big[ \# + \# b^2 q_v^2 + \ldots \big]  q_u^2 q_v^2 h_{(0)vv}
\end{split} 
\ee
where various (scheme-dependent) numerical coefficients are denoted by $\#$. As we mentioned before, in the appendix to this paper we explain why the left-hand side is the expected trace Ward identity for a deformation with a vector operator with scaling dimension $\D = 3$ in $d=2$. The right-hand side is the unintegrated conformal anomaly to first order in the sources.

As explained in detail in the appendix, we can combine the above equations to find that the current $R_i$, defined as:
\be
R_i dx^i = J_{uu} u du + J_{vu} u dv 
\ee
is conserved up to contact terms,
\be
\del^i R_i + \hf \mathcal A = 0\,,
\ee
reflecting the non-relativistic scale invariance of the theory for finite $b$. In fact, we may rewrite \eqref{eq:tracewardid} as:
\be
J_{vu} = \hf  \mathcal A
\ee
which demonstrates that $J_{vu}$ is a trivial operator, \ie all its correlation functions reduce to contact terms. Since $\del^i J_{ij} = 0$, we find that the operator $J_{uu}$ satisfies $\del_v J_{uu} = 0$ and therefore generates an infinite-dimensional Virasoro symmetry for finite $b$ as well. This `left-moving' Virasoro algebra is thus unaffected by the deformation.

\subsection{Callan-Symanzik equation}
The methods for obtaining the Callan-Symanzik equation in the presence of double-trace counter\-terms were extensively discussed in \cite{baltlogs}. We find that it takes the form:
\be
\label{cseqnfinb}
\begin{split} 
&\int d^2 x\, \Big(2 g_{(0)ij} +  2 h_{(0)ij} \Big) \fdel[S_{\text{ren}}]{h_{(0)ij}}   
\\
& + \int d^2 x \,  \Big( 2 b + 2 \big[ 1 + \ldots \big] a_{(0)u} +  \qt b^3 \big[ 1 + \ldots \big] \del_u \del_v h_{(0)vv} +   \# b^4\big[1 + \ldots \big]  \del_u \del_v^3 a_{(0)v}\Big)  \fdel[S_{\text{ren}}]{a_{(0)u}}  \\
&+ \int d^2 x \, \Big(2 \big[ 1 + \ldots \big] a_{(0)v} + \frac{1}{12} b^3 \big[ 1 + \ldots \big] \del_v^2  h_{(0)vv} \Big) \fdel[S_{\text{ren}}]{a_{(0)v}} \\
&+ \int d^2 x \, \Big( - \hf b^2 \big[ 1 + \ldots \big] \tilde R[h_{(0)}]  + b^2  \big[ 1+ \ldots \big] \del_u^2 h_{(0)vv}\\
& \qquad \qquad \qquad \qquad + \# b^3 \big[1 + \ldots \big] \del_u \del_v^3 a_{(0)u} + \# b^3 \big[1 + \ldots \big] \del_u^2 \del_v^2 a_{(0)u} \Big) \fdel[S_{\text{ren}}]{h_{(0)uu}}\\
&+ \int d^2 x \, 2 b^3 \fdel[S_{\text{ren}}]{a_{(0)u}} \fdel[S_{\text{ren}}]{h_{(0)uu}} =  - \mathcal B
\end{split}
\ee
Here the dots represent infinite series in $b^2 q_v^2$ and the $\#$ represent scheme-dependent numbers. The conformal anomaly $\mathcal B$ can be conveniently represented as:
\be
\label{anomalyfinb}
\mathcal B = s^t B s
\ee
with $s$ the vector of local sources,
\be
s^t = \big( a_{(0)u}, a_{(0)v},h_{(0)vv},\tilde R[h_{(0)}] \big)
\ee
and the matrix $B$ given by:
\be
B =
\begin{pmatrix}
\hf \del_u \del_v^3 & \# b^2 \del_u^2 \del_v^4 & - \hf b \del_u^2 \del_v^2 & \hf b \del_v^2\\
0 & \hf \del_u^3 \del_v & - \hf b \del_u^3 \del_v & \# b^3 \del_u \del_v^3\\
0 & 0 & \qt b^2 \del_u^3 \del_v & \# b^4 \del_u \del_v^3\\
0 & 0 & 0 & O(b^6) 
\end{pmatrix}
\ee
where, although we have not written it explicitly here, each term is expected to receive an infinite series of corrections in $b^2 q_v^2$. Notice that all the terms in $B$ have the correct weight and tensor structure so that $\mathcal B$ is a scalar of dimension $2$.

Let us now explain these equations a bit further. First of all, we recall from \cite{baltlogs} that the final term in \eqref{cseqnfinb}, which is quadratic in functional derivatives, represents operator mixing between single- and double-trace operators at large $N$. From this term we conclude that the deforming operator $\op_v$ mixes with the double-trace operator $\op_v T_{vv}$ at order $b^3$. Furthermore, the new coefficients multiplying the other functional derivatives of the on-shell action in \eqref{cseqnfinb} represent the anomalous scaling dimensions of the corresponding operators. We see that all the anomalous scaling dimensions in \eqref{cseqnfinb} involve derivatives and therefore vanish for constant sources. This is in agreement with the all-order analysis of \cite{us,Kraus:2011pf}, although in those papers the (multi-trace) operator mixing was not taken into account. 

Of course, the conformal anomaly and the beta functions involve many terms and there is no clear structural pattern. (The one exception is that $h_{(0)uv}$ and $h_{(0)uu}$ only appear in the linearized curvature $\tilde R[h_{(0)}]$, which is presumably a consequence of our covariant renormalization scheme.)  Such a plethora of terms is actually to be expected, since an irrelevant deformation generally leads to the addition of  all possible counterterms consistent with the residual symmetries. Whenever these counterterms are associated with logarithmic divergences, a corresponding term in the Callan-Symanzik equation appears and this leads to the multitude of terms we observe in the above equations.

At higher orders in $b$ we expect no new tensor structures to appear in the Callan-Symanzik equation. Instead, we expect that all the terms in the equations \eqref{cseqnfinb} and \eqref{anomalyfinb} will get corrected with infinite series in the scalar operator $b^2 q_v^2$. For finite $b$, such series will generally resum to terms that are non-local in the $v$ direction. This non-locality in the Callan-Symanzik equation is an unavoidable consequence of the finite irrelevant deformation. It signals that for finite values of $b$ the $v$-dependence of our results will be scheme-dependent and essentially devoid of any physical information. Fortunately, as explained already in \cite{us}, our deformation retains locality in the $u$ direction. The $u$-dependent part of our results is therefore scheme-independent and representative of the actual dynamics of the deformed field theory.

\subsection{Two-point functions}
\label{subsec:twoptfiniteb}
The two-point functions can be computed by expressing the normalizable modes $a_{(4)i}$ and $h_{(2)ij}$ in terms of the sources $a_{(0)i}$ and $h_{(0)ij}$ by demanding regularity of the solution in the interior of the spacetime. In our case regularity in the interior can be easily imposed by going back to the solution written in terms of Bessel functions \eqref{eq:arbessels} and expanding it around $r = - \infty$. In terms of the integration constants $c_{\pm 1}$ and $c_{\pm 9}$ introduced in \eqref{eq:arbessels} we find that the necessary conditions become $c_{+1} = c_{-1}$ and $c_{+9} = c_{-9}$. In terms of the $\psi_{(0)i}$ and $\psi_{(4)i}$ integration constants we then find that:
\be
\begin{split}
\psi_{(4)u} = \frac{1}{2b^2} \chi_9 \psi_{(0)v}\qquad \qquad
\psi_{(4)v} = \frac{1}{2b^2} \chi_1 q_u q_v \psi_{(0)u}
\end{split}
\ee
where we introduced:
\be
\begin{split}
\chi_9 &= - \frac{2^{2 - s_9} b^2  (3 + s_9) \G(1-s_9)}{\G(2 + s_9)} q_u^2  (q_u q_v - i \e)^{s_9 - 2} 
\\ \chi_1 &= - \frac{2^{- s_1} b^2 \G(1-s_1)}{\G(2 + s_1)} q_v^2 (q_u q_v - i \e)^{s_1 - 1} 
\end{split}
\ee
to simplify the expressions below. These expressions, combined with the expressions \eqref{eq:chi2constraints} which express the $\chi_{(2)ij}$ in terms of the $\chi_{(0)ij}$, can be substituted in our definition for the sources \eqref{eq:redefna0}-\eqref{eq:redefnh2} to express all the normalizable modes in terms of the sources. We can then take a further functional derivative of the one-point functions \eqref{eq:1ptvectorfull} and \eqref{eq:1ptmetricfull} to arrive at the two-point functions given below.

For the two-point functions of the vector field we find it to be advantageous to introduce a new operator $\tilde \op_u$ which will replace $\op_u$, defined as:
\be
\tilde \op_u = \big[ 1 - \frac{1}{72} b^2 q_v^2 - \frac{743}{3456} b^4 q_v^4 + \ldots \big]\Big( \op_u + \frac{1}{3} b^2 q_u q_v \big[1 + b^2 q_v^2 + \ldots \big] \op_v \Big)
\ee
which re-diagonalizes the two-point functions in the $(\op_u,\op_v)$ sector and also introduces a convenient normalization for $\tilde \op_u$. Furthermore, we also define a new normalization for the other operators by introducing:
\be
\hat X = X \big[ 1 - \frac{1}{8} b^2 q_v^2 + \frac{11}{128} b^4 q_v^4 + \ldots \big]  \qquad \qquad X \in \{ \op_v, J_{uu},J_{uv},J_{vu},J_{vv} \}
\ee
where the factor on the right-hand side is the square root of the factor $[1 + b^2 q_v^2 /4 - b^4 q_v^2 / 8 + \ldots]$ appearing in various one-point functions in \eqref{eq:1ptvectorfull} and \eqref{eq:1ptmetricfull}. 

With these conventions we find the following correlation functions for two vector operators: 
\be
\label{eq:vecvecfinb}
\begin{split}
\vev{\tilde \op_u \tilde \op_u} &=  - i \frac{2}{b^2} \int \frac{d^2 q}{(2\pi)^2} e^{i q\cdot x} \big(\chi_9 - \frac{3 q_u^4}{2 (q_u q_v - i \e)} \big)\\ 
\vev{\hat \op_v \hat \op_v} &=  - i \int \frac{d^2 q}{(2\pi)^2} e^{i q\cdot x} \frac{q_u q_v}{2b^2} \frac{2 \chi_1 - 1}{1 - \chi_1} \\
\vev{\tilde \op_u \hat \op_v} &= 0
\end{split}
\ee
where on the right-hand side of these equations we omitted scheme-dependent contact terms. Just as for the Callan-Symanzik equation at finite $b$, these contact terms will generally involve infinite series in $b^2 q_v^2$, so for finite $b$ they become arbitrary functions of $v^2 / b^2$. (Furthermore, analyticity at the origin of momentum space dictates that the corresponding position-space expressions are integrable in the $v$ direction.) This is again precisely the infinite scheme dependence and non-locality typically associated with a finite irrelevant deformation. On the other hand, all of these infinitely many contact terms are at most polynomial in $q_u$ and therefore the position-space expressions will indeed be contact terms in the $u$ direction. 

Two-point functions involving one (new) energy-momentum and one vector operators take the form:
\begin{align}
\vev{\hat J_{uu} \tilde \op_u} &= 0 && \vev{\hat J_{vv} \tilde \op_u}  = 0 \nonumber\\
\vev{\hat J_{uu} \op_v} &= 0 && \vev{\hat J_{vv}  \hat \op_v} = i \pi \int \frac{d^2 q}{(2\pi)^2} \frac{q_v^2}{b} \frac{2 \chi_1 - 1}{1 - \chi_1}
\end{align}
Finally, the two-point functions involving only the energy-momentum tensor become:
\be
\begin{split}
\vev{\hat J_{uu} \hat J_{uu}} &= -  4 \pi^2 i\int \frac{d^2 q}{(2\pi)^2} e^{i q\cdot x} \frac{q_u^4}{(q_u q_v - i \e)} = \frac{12 \pi}{u^4}\\
\vev{\hat J_{uu} \hat J_{uv}} &= 0\\
\vev{\hat J_{uu} \hat J_{vv}} &= 0\\
\vev{\hat J_{uv} \hat J_{uv}} &= - 2 \pi^2 i \int \frac{d^2 q}{(2\pi)^2} e^{i q\cdot x}\frac{q_u q_v}{(1 - \chi_1)}\\
\vev{\hat J_{uv} \hat J_{vv}} &=  2 \pi^2 i  \int \frac{d^2 q}{(2\pi)^2} e^{i q\cdot x}\frac{q_v^2}{(1 - \chi_1)}\\
\vev{\hat J_{vv} \hat J_{vv}} &= - 2 \pi^2 i \int \frac{d^2 q}{(2\pi)^2} e^{i q\cdot x}\frac{q_v^4}{(q_u q_v - i \e) (1 - \chi_1)}
\end{split} 
\ee
where we again suppressed contact terms on the right-hand sides. One may explicitly verify that the Ward identities discussed above are satisfied.

Performing the inverse Fourier transforms in the above correlation functions seems quite complicated, although it is not hard to see that for small $q_u$ and fixed $q_v$, so at large separation in the $u$ direction, all correlation functions reduce to their AdS$_3$ counterparts. We can also try to scale out the $u$-dependence to find a mixed position-momentum space expressions. In this way we find that for example the first equation in \eqref{eq:vecvecfinb} becomes:
\be
\vev{\tilde \op_u (u,q_v) \tilde \op_u(u', - q_v)} \simeq |u - u'|^{- s_9 - 1} - \frac{9}{\pi b^2 (u - u')^4}
\ee 
The last factor can be removed by shifting $\tilde \op_u \to \tilde \op_u + \hf \sqrt{3} \hat J_{uu} / (\pi b)$ although in that case the $b \to 0$ limit of the correlator is no longer smooth. From a non-relativistic viewpoint $q_v$ becomes a quantum number (the particle number in condensed matter applications \cite{Son:2008ye,Balasubramanian:2008dm}) and the powers of $u$ are the non-relativistic dilatation weights. We see that these weights have become dependent on $q_v$, so for an uncompactified $v$ direction the spectrum of the non-relativistic dilatation operator is continuous. For applications to condensed matter physics one should demand a discrete spectrum. One method for obtaining this spectrum is to compactify the lightlike $v$ direction. A discussion of the subtleties associated with this DLCQ can be found in \cite{Maldacena:2008wh}.

\section{Conclusions}
We have holographically computed the two-point functions of the deforming vector operator and the energy-momentum tensor for the field theory dual to the Schr\"odinger backgrounds using a consistent irrelevant deformation picture. We discussed in detail the non-relativistic Ward identities satisfied by the correlation functions. The holomorphicity of the operator $J_{uu}$ in particular indicates that the deformation keeps the `left-moving' symmetries of the original CFT intact, which is in agreement with the viewpoint that the deformation is exactly marginal from a non-relativistic point of view. The right-moving Virasoro is however broken already at first order in $b$. It is also interesting to note that the Callan-Symanzik equation demonstrates mixing between single- and double-trace operators. This leads one to conclude that the question whether the deformation is of single- or multi-trace form becomes renormalization scheme dependent. It would be interesting to see if our results could be matched to computations in dipole theories.

One very valuable general lesson arising from our analysis is that, to any \emph{finite} order in any irrelevant deformation parameter, the sources are \emph{always} the term of order $\exp(-\D_- r)$ in the radial expansion of the fields, with $\D_-$ being the appropriate scaling dimension for the sources in the absence of the deformation. In other words, the location of the sources in the radial expansion of the fields is unmodified to any finite order in perturbation theory. Only when we consider all-order effects in the deformation parameter do we observe a modification of the location of the source terms. Interestingly, in such cases the analysis becomes very subtle as the sources generally do \emph{not} correspond to the leading terms of the radial expansion of the bulk fields. We demonstrated the resolution of these issues explicitly in section \ref{sec:asymptsol}.

We would like to stress that the proper identification of the source terms lies at the heart of the problem of defining a proper holographic dictionary for these non-AlAdS spacetimes. Notice that this is in essence a gravity question, since the correct identification of the sources also determines the proper boundary conditions for the bulk dynamics. This is an essential step in order to for example define conserved charges, the notion of `asymptotically Schr\"odinger spacetimes' and the Schr\"odinger equivalent of `normalizable modes'. In this light it is interesting to note that, if one expresses these proper boundary conditions in terms of the induced bulk fields at finite $r$, the presence of multi-trace counterterms dictates that these boundary conditions will involve the conjugate momenta of the fields as well. Dictating simple falloff conditions for the metric, which happens frequently in for example an asymptotic symmetry group analysis, is therefore not sufficiently general for spacetimes with complicated asymptotics.

Our results come with a significant scheme dependence which indicates a certain dependence on the UV completion of the boundary theory even at the level of classical bulk gravity. It would be interesting to see whether there perhaps exists a `holographic renormalization scheme' which would be natural from the perspective of the gravity dual. In this light it is also important to realize that at large $N$ one can perform additional redefinitions of the sources which correspond to e.g. multi-trace deformations of the theory. We have fixed this freedom by demanding that the sources are the terms at order $\exp(-\D_- r)$ in the radial expansion of the fields, at least to every finite order in perturbation theory. It may however be worthwhile to further explore this freedom. 

Our analysis paves the way for the computation of $n$-point correlation functions in three dimensions. Furthermore, there should be no conceptual difficulties in the extraction of correlation functions from higher-dimensional Schr\"odinger spacetimes or from black holes with Schr\"odinger asymptotics \cite{Adams:2008wt,Herzog:2008wg,Maldacena:2008wh}. It would also be very interesting to generalize our results to other spacetimes which are not asymptotically AdS, for example the Lifshitz spacetimes or the spacetimes of \cite{Hashimoto:1999ut,Maldacena:1999mh} dual to non-commutative gauge theories.

\section*{Acknowledgments}
We would like to thank Marco Baggio, Chris Beem, Geoffrey Comp\`ere, Mukund Rangamani, Leonardo Rastelli and Andrew Strominger for discussions. We would furthermore like to thank Nikolay Bobev, Kostas Skenderis, Marika Taylor and especially Monica Guica for earlier collaboration on this project as well as numerous discussions. The author would like to acknowledge the KITP in Santa Barbara for hospitality during the final stages of this work. This research was supported in part by the National Science Foundation under Grants PHY-0969739 and PHY-0551164. 

\appendix

\section{Vector deformations in quantum field theory}
Consider the partition function $Z$ of a $d$-dimensional conformal field theory in a background metric $g_{ij}$ deformed by a source $b_i$ for a vector operator $\op_i$ with dimension $\D$. In path-integral notation the partition function can be abstractly written as:
\be
Z[g_{ij}, b_i] = \int  D\Phi \exp \Big( - i S[g_{ij}, \Phi] - i \int d^d x \sqrt{g} g^{ij} b_i \op_j  + (i \e \text{ insertions}) \Big)\,.
\ee
In this appendix we discuss the modification of the field theory Ward identities that this deformation brings about. The $i \e$ insertions are important for obtaining the correct correlation functions in Lorentzian signature but we will ignore them here. 

In the conventions imposed by the holographic analysis the energy-momentum tensor is defined to be the operator dual to the boundary metric, more precisely
\be
T^{ij} = - \frac{4 \pi i}{\sqrt{-g}} \fdel{g_{ij}} Z[g_{ij},b_i] = \hat T^{ij} - 2 \pi (g^{ij} b^k \op_k -  b^i \op^j -  b^j \op^i)
\ee
with $\hat T^{ij}$ the energy-momentum tensor before the deformation. Notice that we hold the one-form $b_i$ fixed while varying the metric.

Under a diffeomorphism along a vector field $\z^i$ both the metric $g_{ij}$ and the vector field $b_i$ transform by their Lie derivatives:
\be
\delta g_{ij} = \cdel_i \z_j + \cdel_j \z_i \qquad \qquad \delta b_i = \z^j \cdel_j b_i + b_j \cdel_i \z^j 
\ee    
and in the absence of a diffeomorphism anomaly in the path integral measure we should find that the partition function is invariant,
\be
\delta_\z Z = 0\,,
\ee
leading to
\be
\cdel^i (T_{ij} - 2 \pi \op_i b_j) + 2 \pi (\cdel_j b_i) \op^i = 0\,.    
\ee
We will be working in flat space and with a constant $b$, so $\cdel_j b_i = 0$. In that case we find that the conserved current associated to translation invariance is given by:
\be
\label{eq:diffwi}
J_{ij} \equiv T_{ij} - 2 \pi \op_i b_j \qquad \qquad \del^i J_{ij} = 0
\ee
This is indeed confirmed by the holographic analysis in the main text.

Since the new energy-momentum tensor $J_{ij}$ is not symmetric, Lorentz invariance is manifestly broken, even if $\del_i b_j = 0$. For example in $d=2$ a boost can be implemented by a diffeomorphism with $\z^j = \e^j_{\ph{j}k}x^k$. In that case we find that
\be
\label{eq:boostwi}
\del^i (T_{ij}\e^j_{\ph{j}k} x^k - 2 \pi  b_j\e^j_{\ph{j}k} x^k \op_i) + 2 \pi b_j\e^j_{\ph{j}k} \op^k = 0\,,
\ee 
which does not  integrate to a conservation equation.

Let us now discuss Weyl covariance to first order in $b$. A local scale transformation is implemented by sending $g_{ij} \to 2 \lambda g_{ij}$ and $b_i \to \lambda ( \D - d + 1) b_i$ with $\lambda$ the infinitesimal scale transformation parameter. Assuming that the transformation of the path integral measure results in a local conformal anomaly $\mathcal A$, the associated Ward identity would become:
\be
\label{eq:tracewi}
T_i^i - 2 \pi (\D - d + 1) b_i \op^i + \mathcal A[g_{ij}] = 0\,.
\ee
This equation is valid to first order in $b$. At higher orders one generally finds that the classical scaling dimensions get corrections, so new one-point functions may appear and the classical scaling dimension $\D$ becomes a matrix of anomalous dimensions. Similarly, the conformal anomaly $\mathcal A$ may receive $b$-dependent corrections.

We may combine \eqref{eq:tracewi} with the diffeomorphism Ward identity \eqref{eq:diffwi} to find that:
\be
\label{eq:dilwirewr}
\del^i (T_{ij} x^j  - 2 \pi x_j b^j \op_i) - 2 \pi (\D - d) b_i \op^i  + \mathcal A[g_{ij}] = 0\,. 
\ee
Now let us specialize to the case where $d=2$ and $b_i$ is null, $b_i dx^i = b du$. For such a vector one verifies that $b_j \e^{j}_{\ph{j}k} = - b_k$ (in conventions where $\e_{uv} = +1$) and therefore we can combine \eqref{eq:dilwirewr} and \eqref{eq:boostwi} to find that:
\be
\del^i R_i + \hf \mathcal A[g_{ij}] = 0 \,,
\ee
with
\be
R_i = T_{ij} (x^j - (\D - d) \e^j_{\ph{j}k} x^k)  - 2 \pi b_j x^j (\D - d + 1) \op_i\,,
\ee
which is the current associated to the non-relativistic scale invariance of the theory. In the main text we have $\D = 3$ and $d=2$. In that case the current $R_i$ becomes:
\be
R_u =  J_{uu} u \qquad \qquad 
R_v =  J_{vu} u   
\ee
In fact, precisely for these values for $\D$ and $d$ we see from equation \eqref{eq:tracewi} that
\be
J_{vu} = - \hf \mathcal A[g_{ij}]
\ee
and $J_{vu}$ is therefore `trivial' in the sense that its correlation functions reduce to contact terms. The conservation equation $\del^i J_{iu}$ therefore dictates that,
\be
\del_v J_{uu} = 0
\ee
up to contact terms. We thus immediately see that we can build infinitely many conserved currents of the form:
\be
J_{uu} u^k
\ee
for some integer $k$. In other words, the left-moving Virasoro algebra is left intact by the deformation.

\bibliographystyle{utphys}
\bibliography{biblio}

\end{document}

%% file: commands.tex
\usepackage{fullpage}

\usepackage{color}
\usepackage{amsmath, amsfonts, amsthm, amssymb, graphicx}
\usepackage{xspace}
\usepackage{epsfig}
\usepackage{psfrag}

\newcommand{\op}{\ensuremath{\mathcal{O}}\xspace}
\newcommand{\del}{\partial} 
\newcommand{\cdel}{\nabla}
\newcommand{\fdel}[2][]{\ensuremath{\frac{\delta #1}{\delta #2}}}

\newcommand{\vev}[1]{\ensuremath{\langle #1 \rangle}\xspace}
\newcommand{\ph}[1]{\phantom{#1}}

\def\ie{{\it i.e.\ }}

\newcommand{\du}[2]{_{ #1 }^{\phantom{ #1 } #2 }}

\newcommand{\hf}{\frac{1}{2}}
\newcommand{\qt}{\frac{1}{4}}

  \let\g=\gamma \let\d=\delta \let\e=\epsilon
\let\z=\zeta    
 \let\m=\mu \let\n=\nu  \let\r=\rho
\let\s=\sigma    \let\c=\chi 
  \let\D=\Delta  
    
    \let\G=\Gamma

\newcommand{\be}{\begin{equation}}
\newcommand{\ee}{\end{equation}}
\def\ba{\begin{array}}
\def\ea{\end{array}}
\newcommand{\bea}{\begin{eqnarray}}
\newcommand{\eea}{\end{eqnarray}}

%% file: document.bbl
\providecommand{\href}[2]{#2}\begingroup\raggedright\begin{thebibliography}{10}

\bibitem{Son:2008ye}
D.~T. Son, ``{Toward an AdS/cold atoms correspondence: a geometric realization
  of the Schroedinger symmetry},''
  \href{http://dx.doi.org/10.1103/PhysRevD.78.046003}{{\em Phys. Rev.} {\bf
  D78} (2008)  046003},
\href{http://arxiv.org/abs/0804.3972}{{\tt arXiv:0804.3972 [hep-th]}}.

\bibitem{Balasubramanian:2008dm}
K.~Balasubramanian and J.~McGreevy, ``{Gravity duals for non-relativistic
  CFTs},'' \href{http://dx.doi.org/10.1103/PhysRevLett.101.061601}{{\em Phys.
  Rev. Lett.} {\bf 101} (2008)  061601},
\href{http://arxiv.org/abs/0804.4053}{{\tt arXiv:0804.4053 [hep-th]}}.

\bibitem{FeffermanGraham}
C.~Fefferman and C.~Graham, ``{Conformal Invariants},'' {\em Elie Cartan et les
  Math{\'e}matiques d'aujourd'hui (Asterisque 95)} (1985)  .

\bibitem{Graham:1999jg}
C.~R. Graham, ``{Volume and area renormalizations for conformally compact
  Einstein metrics},''
\href{http://arxiv.org/abs/math/9909042}{{\tt arXiv:math/9909042}}.

\bibitem{us}
M.~Guica, K.~Skenderis, M.~Taylor, and B.~C. van Rees, ``{Holography for
  Schrodinger backgrounds},''
\href{http://arxiv.org/abs/1008.1991}{{\tt arXiv:1008.1991 [hep-th]}}.

\bibitem{Costa:2010cn}
R.~Costa and M.~Taylor, ``{Holography for chiral scale-invariant models},''
  \href{http://arxiv.org/abs/1010.4800}{{\tt arXiv:1010.4800 [hep-th]}}.

\bibitem{Kraus:2011pf}
P.~Kraus and E.~Perlmutter, ``{Universality and exactness of Schrodinger
  geometries in string and M-theory},''
  \href{http://dx.doi.org/10.1007/JHEP05(2011)045}{{\em JHEP} {\bf 1105} (2011)
   045},
\href{http://arxiv.org/abs/1102.1727}{{\tt arXiv:1102.1727 [hep-th]}}.

\bibitem{Heemskerk:2010hk}
I.~Heemskerk and J.~Polchinski, ``{Holographic and Wilsonian Renormalization
  Groups},'' \href{http://arxiv.org/abs/1010.1264}{{\tt arXiv:1010.1264
  [hep-th]}}.

\bibitem{Faulkner:2010jy}
T.~Faulkner, H.~Liu, and M.~Rangamani, ``{Integrating out geometry: Holographic
  Wilsonian RG and the membrane paradigm},''
  \href{http://dx.doi.org/10.1007/JHEP08(2011)051}{{\em JHEP} {\bf 1108} (2011)
   051},
\href{http://arxiv.org/abs/1010.4036}{{\tt arXiv:1010.4036 [hep-th]}}.

\bibitem{Brattan:2011my}
D.~Brattan, J.~Camps, R.~Loganayagam, and M.~Rangamani, ``{CFT dual of the AdS
  Dirichlet problem : Fluid/Gravity on cut-off surfaces},''
  \href{http://dx.doi.org/10.1007/JHEP12(2011)090}{{\em JHEP} {\bf 1112} (2011)
   090},
\href{http://arxiv.org/abs/1106.2577}{{\tt arXiv:1106.2577 [hep-th]}}.

\bibitem{Compere:2009qm}
G.~Compere, S.~de~Buyl, S.~Detournay, and K.~Yoshida, ``{Asymptotic symmetries
  of Schrodinger spacetimes},''
  \href{http://dx.doi.org/10.1088/1126-6708/2009/10/032}{{\em JHEP} {\bf 0910}
  (2009)  032},
\href{http://arxiv.org/abs/0908.1402}{{\tt arXiv:0908.1402 [hep-th]}}.

\bibitem{Anninos:2010pm}
D.~Anninos, G.~Compere, S.~de~Buyl, S.~Detournay, and M.~Guica, ``{The Curious
  Case of Null Warped Space},''
  \href{http://dx.doi.org/10.1007/JHEP11(2010)119}{{\em JHEP} {\bf 1011} (2010)
   119},
\href{http://arxiv.org/abs/1005.4072}{{\tt arXiv:1005.4072 [hep-th]}}.

\bibitem{Guica:2011ia}
M.~Guica, ``{A Fefferman-Graham-Like Expansion for Null Warped AdS(3)},''
\href{http://arxiv.org/abs/1111.6978}{{\tt arXiv:1111.6978 [hep-th]}}.

\bibitem{ElShowk:2011cm}
S.~El-Showk and M.~Guica, ``{Kerr/CFT, dipole theories and nonrelativistic
  CFTs},''
\href{http://arxiv.org/abs/1108.6091}{{\tt arXiv:1108.6091 [hep-th]}}.

\bibitem{Song:2011sr}
W.~Song and A.~Strominger, ``{Warped AdS3/Dipole-CFT Duality},''
\href{http://arxiv.org/abs/1109.0544}{{\tt arXiv:1109.0544 [hep-th]}}.

\bibitem{Hofman:2011zj}
D.~M. Hofman and A.~Strominger, ``{Chiral Scale and Conformal Invariance in 2D
  Quantum Field Theory},'' {\em Phys.Rev.Lett.} {\bf 107} (2011)  161601,
\href{http://arxiv.org/abs/1107.2917}{{\tt arXiv:1107.2917 [hep-th]}}.

\bibitem{Wiseman:2008qa}
T.~Wiseman and B.~Withers, ``{Holographic renormalization for coincident
  Dp-branes},'' \href{http://dx.doi.org/10.1088/1126-6708/2008/10/037}{{\em
  JHEP} {\bf 0810} (2008)  037},
\href{http://arxiv.org/abs/0807.0755}{{\tt arXiv:0807.0755 [hep-th]}}.

\bibitem{Kanitscheider:2008kd}
I.~Kanitscheider, K.~Skenderis, and M.~Taylor, ``{Precision holography for
  non-conformal branes},''
  \href{http://dx.doi.org/10.1088/1126-6708/2008/09/094}{{\em JHEP} {\bf 09}
  (2008)  094},
\href{http://arxiv.org/abs/0807.3324}{{\tt arXiv:0807.3324 [hep-th]}}.

\bibitem{Kanitscheider:2009as}
I.~Kanitscheider and K.~Skenderis, ``{Universal hydrodynamics of non-conformal
  branes},'' \href{http://dx.doi.org/10.1088/1126-6708/2009/04/062}{{\em JHEP}
  {\bf 0904} (2009)  062}, \href{http://arxiv.org/abs/0901.1487}{{\tt
  arXiv:0901.1487 [hep-th]}}.

\bibitem{Gouteraux:2011qh}
B.~Gouteraux, J.~Smolic, M.~Smolic, K.~Skenderis, and M.~Taylor, ``{Holography
  for Einstein-Maxwell-dilaton theories from generalized dimensional
  reduction},'' {\em JHEP} {\bf 1201} (2012)  089,
\href{http://arxiv.org/abs/1110.2320}{{\tt arXiv:1110.2320 [hep-th]}}.

\bibitem{Papadimitriou:2011qb}
I.~Papadimitriou, ``{Holographic Renormalization of general dilaton-axion
  gravity},'' {\em JHEP} {\bf 1108} (2011)  119,
\href{http://arxiv.org/abs/1106.4826}{{\tt arXiv:1106.4826 [hep-th]}}.

\bibitem{Aharony:2005zr}
O.~Aharony, A.~Buchel, and A.~Yarom, ``{Holographic renormalization of
  cascading gauge theories},''
  \href{http://dx.doi.org/10.1103/PhysRevD.72.066003}{{\em Phys.Rev.} {\bf D72}
  (2005)  066003},
\href{http://arxiv.org/abs/hep-th/0506002}{{\tt arXiv:hep-th/0506002
  [hep-th]}}.

\bibitem{Borodatchenkova:2008fw}
N.~Borodatchenkova, M.~Haack, and W.~Muck, ``{Towards Holographic
  Renormalization of Fake Supergravity},''
  \href{http://dx.doi.org/10.1016/j.nuclphysb.2009.02.018}{{\em Nucl.Phys.}
  {\bf B815} (2009)  215--239},
\href{http://arxiv.org/abs/0811.3191}{{\tt arXiv:0811.3191 [hep-th]}}.

\bibitem{Papadimitriou:2010as}
I.~Papadimitriou, ``{Holographic renormalization as a canonical
  transformation},''
\href{http://arxiv.org/abs/1007.4592}{{\tt arXiv:1007.4592 [hep-th]}}.

\bibitem{Ross:2009ar}
S.~F. Ross and O.~Saremi, ``{Holographic stress tensor for non-relativistic
  theories},'' \href{http://dx.doi.org/10.1088/1126-6708/2009/09/009}{{\em
  JHEP} {\bf 0909} (2009)  009},
\href{http://arxiv.org/abs/0907.1846}{{\tt arXiv:0907.1846 [hep-th]}}.

\bibitem{Ross:2011gu}
S.~F. Ross, ``{Holography for asymptotically locally Lifshitz spacetimes},''
  \href{http://dx.doi.org/10.1088/0264-9381/28/21/215019}{{\em
  Class.Quant.Grav.} {\bf 28} (2011)  215019},
\href{http://arxiv.org/abs/1107.4451}{{\tt arXiv:1107.4451 [hep-th]}}.

\bibitem{Baggio:2011cp}
M.~Baggio, J.~de~Boer, and K.~Holsheimer, ``{Hamilton-Jacobi Renormalization
  for Lifshitz Spacetime},'' {\em JHEP} {\bf 1201} (2012)  058,
\href{http://arxiv.org/abs/1107.5562}{{\tt arXiv:1107.5562 [hep-th]}}.

\bibitem{Mann:2011hg}
R.~B. Mann and R.~McNees, ``{Holographic Renormalization for Asymptotically
  Lifshitz Spacetimes},'' {\em JHEP} {\bf 1110} (2011)  129,
\href{http://arxiv.org/abs/1107.5792}{{\tt arXiv:1107.5792 [hep-th]}}.

\bibitem{Griffin:2011xs}
T.~Griffin, P.~Horava, and C.~M. Melby-Thompson, ``{Conformal Lifshitz Gravity
  from Holography},''
\href{http://arxiv.org/abs/1112.5660}{{\tt arXiv:1112.5660 [hep-th]}}.

\bibitem{Baggio:2011ha}
M.~Baggio, J.~de~Boer, and K.~Holsheimer, ``{Anomalous Breaking of Anisotropic
  Scaling Symmetry in the Quantum Lifshitz Model},''
\href{http://arxiv.org/abs/1112.6416}{{\tt arXiv:1112.6416 [hep-th]}}.

\bibitem{Kachru:2008yh}
S.~Kachru, X.~Liu, and M.~Mulligan, ``{Gravity Duals of Lifshitz-like Fixed
  Points},'' \href{http://dx.doi.org/10.1103/PhysRevD.78.106005}{{\em Phys.
  Rev.} {\bf D78} (2008)  106005},
\href{http://arxiv.org/abs/0808.1725}{{\tt arXiv:0808.1725 [hep-th]}}.

\bibitem{Taylor:2008tg}
M.~Taylor, ``{Non-relativistic holography},''
\href{http://arxiv.org/abs/0812.0530}{{\tt arXiv:0812.0530 [hep-th]}}.

\bibitem{Hartong:2010ec}
J.~Hartong and B.~Rollier, ``{Asymptotically Schroedinger Space-Times: TsT
  Transformations and Thermodynamics},''
  \href{http://dx.doi.org/10.1007/JHEP01(2011)084}{{\em JHEP} {\bf 1101} (2011)
   084},
\href{http://arxiv.org/abs/1009.4997}{{\tt arXiv:1009.4997 [hep-th]}}.

\bibitem{Adams:2008wt}
A.~Adams, K.~Balasubramanian, and J.~McGreevy, ``{Hot Spacetimes for Cold
  Atoms},'' \href{http://dx.doi.org/10.1088/1126-6708/2008/11/059}{{\em JHEP}
  {\bf 0811} (2008)  059},
\href{http://arxiv.org/abs/0807.1111}{{\tt arXiv:0807.1111 [hep-th]}}.

\bibitem{Herzog:2008wg}
C.~P. Herzog, M.~Rangamani, and S.~F. Ross, ``{Heating up Galilean
  holography},'' \href{http://dx.doi.org/10.1088/1126-6708/2008/11/080}{{\em
  JHEP} {\bf 0811} (2008)  080},
\href{http://arxiv.org/abs/0807.1099}{{\tt arXiv:0807.1099 [hep-th]}}.

\bibitem{Maldacena:2008wh}
J.~Maldacena, D.~Martelli, and Y.~Tachikawa, ``{Comments on string theory
  backgrounds with non- relativistic conformal symmetry},''
  \href{http://dx.doi.org/10.1088/1126-6708/2008/10/072}{{\em JHEP} {\bf 10}
  (2008)  072},
\href{http://arxiv.org/abs/0807.1100}{{\tt arXiv:0807.1100 [hep-th]}}.

\bibitem{Bergman:2001rw}
A.~Bergman, K.~Dasgupta, O.~J. Ganor, J.~L. Karczmarek, and G.~Rajesh,
  ``{Nonlocal field theories and their gravity duals},''
  \href{http://dx.doi.org/10.1103/PhysRevD.65.066005}{{\em Phys. Rev.} {\bf
  D65} (2002)  066005},
\href{http://arxiv.org/abs/hep-th/0103090}{{\tt arXiv:hep-th/0103090}}.

\bibitem{Bobev:2011qx}
N.~Bobev and B.~C. van Rees, ``{Schrodinger Deformations of $AdS_3 x S^3$},''
  \href{http://dx.doi.org/10.1007/JHEP08(2011)062}{{\em JHEP} {\bf 1108} (2011)
   062},
\href{http://arxiv.org/abs/1102.2877}{{\tt arXiv:1102.2877 [hep-th]}}.

\bibitem{balt}
B.~C. van Rees, ``{Holographic renormalization for irrelevant operators and
  multi-trace counterterms},'' \href{http://arxiv.org/abs/1102.2239}{{\tt
  arXiv:1102.2239 [hep-th]}}.

\bibitem{baltlogs}
B.~C. van Rees, ``{Irrelevant deformations and the holographic Callan- Symanzik
  equation},''
\href{http://arxiv.org/abs/1105.5396}{{\tt arXiv:1105.5396 [hep-th]}}.

\bibitem{deHaro:2000xn}
S.~de~Haro, S.~N. Solodukhin, and K.~Skenderis, ``{Holographic reconstruction
  of spacetime and renormalization in the AdS/CFT correspondence},''
  \href{http://dx.doi.org/10.1007/s002200100381}{{\em Commun. Math. Phys.} {\bf
  217} (2001)  595--622},
\href{http://arxiv.org/abs/hep-th/0002230}{{\tt arXiv:hep-th/0002230}}.

\bibitem{Skenderis:1999nb}
K.~Skenderis and S.~N. Solodukhin, ``Quantum effective action from the
  {AdS/CFT} correspondence,'' {\em Phys. Lett.} {\bf B472} (2000)  316--322,
\href{http://arxiv.org/abs/hep-th/9910023}{{\tt hep-th/9910023}}.

\bibitem{Bianchi:2001kw}
M.~Bianchi, D.~Z. Freedman, and K.~Skenderis, ``{Holographic
  renormalization},'' {\em Nucl. Phys.} {\bf B631} (2002)  159--194,
\href{http://arxiv.org/abs/hep-th/0112119}{{\tt arXiv:hep-th/0112119}}.

\bibitem{Skenderis:2002wp}
K.~Skenderis, ``Lecture notes on holographic renormalization,'' {\em Class.
  Quant. Grav.} {\bf 19} (2002)  5849--5876,
\href{http://arxiv.org/abs/hep-th/0209067}{{\tt hep-th/0209067}}.

\bibitem{Papadimitriou:2004ap}
I.~Papadimitriou and K.~Skenderis, ``{AdS / CFT correspondence and geometry},''
  {\em Proceedings Strasbourg 2003, AdS/CFT correspondence, ed. O. Biquard}
  73,
\href{http://arxiv.org/abs/hep-th/0404176}{{\tt hep-th/0404176}}.

\bibitem{Brown:1986nw}
J.~D. Brown and M.~Henneaux, ``{Central Charges in the Canonical Realization of
  Asymptotic Symmetries: An Example from Three-Dimensional Gravity},''
\href{http://dx.doi.org/10.1007/BF01211590}{{\em Commun. Math. Phys.} {\bf 104}
  (1986)  207--226}.

\bibitem{Henningson:1998gx}
M.~Henningson and K.~Skenderis, ``{The holographic Weyl anomaly},'' {\em JHEP}
  {\bf 07} (1998)  023,
\href{http://arxiv.org/abs/hep-th/9806087}{{\tt arXiv:hep-th/9806087}}.

\bibitem{Henningson:1998ey}
M.~Henningson and K.~Skenderis, ``{Holography and the Weyl anomaly},'' {\em
  Fortsch. Phys.} {\bf 48} (2000)  125--128,
\href{http://arxiv.org/abs/hep-th/9812032}{{\tt arXiv:hep-th/9812032}}.

\bibitem{Skenderis:2009nt}
K.~Skenderis, M.~Taylor, and B.~C. van Rees, ``{Topologically Massive Gravity
  and the AdS/CFT Correspondence},''
  \href{http://dx.doi.org/10.1088/1126-6708/2009/09/045}{{\em JHEP} {\bf 09}
  (2009)  045},
\href{http://arxiv.org/abs/0906.4926}{{\tt arXiv:0906.4926 [hep-th]}}.

\bibitem{Skenderis:2008dh}
K.~Skenderis and B.~C. van Rees, ``{Real-time gauge/gravity duality},''
  \href{http://dx.doi.org/10.1103/PhysRevLett.101.081601}{{\em Phys.Rev.Lett.}
  {\bf 101} (2008)  081601},
\href{http://arxiv.org/abs/0805.0150}{{\tt arXiv:0805.0150 [hep-th]}}.

\bibitem{Skenderis:2008dg}
K.~Skenderis and B.~C. van Rees, ``{Real-time gauge/gravity duality:
  Prescription, Renormalization and Examples},''
\href{http://arxiv.org/abs/0812.2909}{{\tt arXiv:0812.2909 [hep-th]}}.

\bibitem{Leigh:2009eb}
R.~G. Leigh and N.~N. Hoang, ``{Real-Time Correlators and Non-Relativistic
  Holography},'' \href{http://dx.doi.org/10.1088/1126-6708/2009/11/010}{{\em
  JHEP} {\bf 0911} (2009)  010},
\href{http://arxiv.org/abs/0904.4270}{{\tt arXiv:0904.4270 [hep-th]}}.

\bibitem{Papadimitriou:2005ii}
I.~Papadimitriou and K.~Skenderis, ``{Thermodynamics of asymptotically locally
  AdS spacetimes},'' {\em JHEP} {\bf 08} (2005)  004,
\href{http://arxiv.org/abs/hep-th/0505190}{{\tt hep-th/0505190}}.

\bibitem{Klebanov:1999tb}
I.~R. Klebanov and E.~Witten, ``{AdS/CFT correspondence and symmetry
  breaking},'' \href{http://dx.doi.org/10.1016/S0550-3213(99)00387-9}{{\em
  Nucl. Phys.} {\bf B556} (1999)  89--114},
\href{http://arxiv.org/abs/hep-th/9905104}{{\tt arXiv:hep-th/9905104}}.

\bibitem{Papadimitriou:2007sj}
I.~Papadimitriou, ``{Multi-Trace Deformations in AdS/CFT: Exploring the Vacuum
  Structure of the Deformed CFT},''
  \href{http://dx.doi.org/10.1088/1126-6708/2007/05/075}{{\em JHEP} {\bf 0705}
  (2007)  075}, \href{http://arxiv.org/abs/hep-th/0703152}{{\tt
  arXiv:hep-th/0703152 [hep-th]}}.

\bibitem{Berkooz:2002ug}
M.~Berkooz, A.~Sever, and A.~Shomer, ``{Double-trace deformations, boundary
  conditions and spacetime singularities},'' {\em JHEP} {\bf 05} (2002)  034,
\href{http://arxiv.org/abs/hep-th/0112264}{{\tt arXiv:hep-th/0112264}}.

\bibitem{Witten:2001ua}
E.~Witten, ``{Multitrace operators, boundary conditions, and AdS / CFT
  correspondence},'' \href{http://arxiv.org/abs/hep-th/0112258}{{\tt
  arXiv:hep-th/0112258 [hep-th]}}.

\bibitem{Elitzur:2005kz}
S.~Elitzur, A.~Giveon, M.~Porrati, and E.~Rabinovici, ``{Multitrace
  deformations of vector and adjoint theories and their holographic duals},''
  \href{http://dx.doi.org/10.1088/1126-6708/2006/02/006}{{\em JHEP} {\bf 02}
  (2006)  006},
\href{http://arxiv.org/abs/hep-th/0511061}{{\tt arXiv:hep-th/0511061}}.

\bibitem{Mueck:2002gm}
W.~Mueck, ``{An improved correspondence formula for AdS/CFT with multi- trace
  operators},'' \href{http://dx.doi.org/10.1016/S0370-2693(02)01487-9}{{\em
  Phys. Lett.} {\bf B531} (2002)  301--304},
\href{http://arxiv.org/abs/hep-th/0201100}{{\tt arXiv:hep-th/0201100}}.

\bibitem{Sever:2002fk}
A.~Sever and A.~Shomer, ``{A note on multi-trace deformations and AdS/CFT},''
  {\em JHEP} {\bf 07} (2002)  027,
\href{http://arxiv.org/abs/hep-th/0203168}{{\tt arXiv:hep-th/0203168}}.

\bibitem{Akhmedov:2002gq}
E.~T. Akhmedov, ``{Notes on multi-trace operators and holographic
  renormalization group},''
\href{http://arxiv.org/abs/hep-th/0202055}{{\tt arXiv:hep-th/0202055}}.

\bibitem{Nakayama:2011fe}
Y.~Nakayama, ``{Gravity Dual for Hofman-Strominger Theorem},''
\href{http://arxiv.org/abs/1112.0635}{{\tt arXiv:1112.0635 [hep-th]}}.

\bibitem{Hashimoto:1999ut}
A.~Hashimoto and N.~Itzhaki, ``{Non-commutative Yang-Mills and the AdS/CFT
  correspondence},''
  \href{http://dx.doi.org/10.1016/S0370-2693(99)01037-0}{{\em Phys. Lett.} {\bf
  B465} (1999)  142--147},
\href{http://arxiv.org/abs/hep-th/9907166}{{\tt arXiv:hep-th/9907166}}.

\bibitem{Maldacena:1999mh}
J.~M. Maldacena and J.~G. Russo, ``{Large N limit of non-commutative gauge
  theories},'' {\em JHEP} {\bf 09} (1999)  025,
\href{http://arxiv.org/abs/hep-th/9908134}{{\tt arXiv:hep-th/9908134}}.

\end{thebibliography}\endgroup
